\documentclass[%
 aip,
rsi,%
 amsmath,amssymb,
 reprint,%
nofootinbib
]{revtex4-1}
\usepackage{pgfplots}
\usepackage{listings} % was minted
\usepackage{breqn}
\usepackage{braket}
\usepackage{makecell}
\usepackage{dcolumn}
\usepackage{bm}
\usepackage{grffile}
\usepackage{hyperref}
\usepackage{graphicx}
\usepackage{verbatim}
\usepackage{calligra}
\DeclareMathAlphabet{\mathcalligra}{T1}{calligra}{m}{n}
\DeclareFontShape{T1}{calligra}{m}{n}{<->s*[2.2]callig15}{}

\usepackage{amsmath,empheq,amsfonts,amsthm,bm}
\hypersetup{
    colorlinks=true,
    linkcolor=blue,
    filecolor=magenta,      
    urlcolor=cyan,
}
\newcommand{\fig}[2]{
\begin{figure}[H]
\centering
\includegraphics[width=0.50\textwidth]{images/#1}
\caption{#2}
\end{figure}
}
\usepackage{float}
\lstdefinestyle{mystyle}{
    basicstyle=\ttfamily\footnotesize,
    breakatwhitespace=false,         
    breaklines=true,                 
    captionpos=b,                    
    keepspaces=true,                 
    numbersep=2pt,                  
    showspaces=false,                
    showstringspaces=false,
    showtabs=false,   
    tabsize=2
}
\DeclareUnicodeCharacter{2212}{-}

\begin{document}
\preprint{AIP/123-QED}

\title[On the Behavior of Superconductors of High Critical Temperatures Outside Schwarzchild Black Holes in AdS Space]{On the Behavior of Superconductors of High Critical Temperatures Outside Schwarzchild Black Holes in AdS Space}%{A Thermal and Structural Analysis of Biomaterials in the Qinghai-Tibetan Plateau} % Force line breaks with \\

\author{D. R. Musk}
 \altaffiliation{Stanford University Online High School}
 % With Professor Frank Peiris
 % Lines break automatically or can be forced with \\

\date{\today}% It is always \today, today,

\begin{abstract}
    The physical analysis of condensed matter systems can be difficult due to strong coupling, but the mathematical context of the AdS/CFT correspondence enables non-perturbative descriptions in terms of dual weakly coupled systems. This brief review explores the holographic condensed matter applications of AdS/CFT, particularly through the lens of a high-$T_c$ superconductor outside a Schwarzchild black hole in Anti-de Sitter space. A simple two-dimensional electron condensate Lagrangian is examined first, as employed by G. T. Horowitz, later used to calculate a frequency-dependent conductivity and a free energy analysis; the asymptotics of both in this procedure, as examined by P. Säterskog, are also reviewed. An extended Lagrangian with a higher order Maxwell term is assessed thereafter, with a conductivity peak obtained at low frequencies described well by the Drude model in certain limits. The behavior of Drude model parameters in these limits is also investigated.
\end{abstract}
                             
           % The (very tentative) paper keywords
\maketitle
\section{\label{sec:level1}Introduction to the Correspondence in Classical Bulk Theory}
\ \ \ \ The AdS/CFT correspondence was first conjectured in 1997 by Juan Maldacena, relating the physics of a string theory in Anti-de Sitter space (AdS) to a conformal field theory (CFT) on the boundary of a AdS space, a maximally symmetric space of constant negative curvature. There is currently no proof of the correspondence, although it has been extensively tested. The strength of the duality is predominantly due to the bulk theory being weakly coupled when the boundary theory is strongly coupled and vice versa. This lets us solve otherwise computationally intractable problems on the strongly coupled side by solving them on the weakly coupled side.

Examples of strongly coupled systems exhibiting critical behavior (that are thereby also applications of the AdS/CFT correspondence) are quark-gluon plasma\cite{Janik2006}, high-$T_c$, high-$T_c$ superconductors\cite{Hartnoll2009}, and possibly graphene\cite{Hartnoll2009}. The focus of the present article, however, is on that middle item: superconductors with high critical temperatures. These types of superconductors are generally layered and electrons within them move in two dimensions. 

There is currently no widely accepted theory describing high-$T_c$ superconductors possibly due to the strong coupling that makes such a theoretical understanding rather difficult\cite{Hartnoll2009}. These devices may be in the vicinity of a quantum critical point\cite{Broun2008} and therefore exhibit scale invariance, further motivating the use of a conformal field theory. 

Mathematically, the correspondence can be formulated through \cite{hartnoll8}
\begin{align}
    Z_{\text{bulk}} (\delta \psi_{(0)}) = \braket{\exp(i \int{\text{d}^d x \sqrt{g_0} \delta \psi_{(0)} \mathcal{O})}}_{\text{CFT}},
    \label{eqn:bulk_partition}
\end{align}
for dual function $\mathcal{O}$ and partition function for the bulk theory with boundary condition $\psi_{(0)}$ at the conformal boundary $Z_{bulk}(\psi_{(0)})$, with boundary background field $\psi_{(0)}$ as source of the operator $\mathcal{O}$ such that
\begin{align}
    \mathcal{O} = \frac{\delta S_{\text{CFT}}}{\delta \psi_{(0)}},
    \label{eqn:expectation}
\end{align}
with CFT action $S_{\text{CFT}}$. Overall, the expectation value of Equation $~\ref{eqn:bulk_partition}$ is then of a field theory at a temperature given by the Euclidean time-periodicity of the path integral for the partition function.

The bulk theory becomes classical for a boundary gauge theory with a large number of colors, a large $N$. This analysis will not involve a large-$N$ theory, but a similar effect is expected for certain strongly coupled boundary theories\cite{McGreevy2010}, thus one can be assumed for this paper.

Given some source $J^{\alpha}$ for a vector field $A_a$ such that
\begin{align}
    J^\alpha = \frac{\delta S_{\text{CFT}}}{\delta A_{(0)a}},
\end{align}
we may determine CFT expectation values from the classical bulk
\begin{align}
    \braket{J(x)}_{\text{CFT}} = \frac{\delta S_c(\psi_0)}{\delta A_{(0)a}(x)}|_{A_{(0)a=0}}.
\end{align}
The functional derivative needed to calculate these expectation values is the change in total on-shell action when the boundary value of the source field is changed. This can in the case of the operator $\mathcal{O}$ and the source $\psi$ be calculated through the following procedure:
\begin{enumerate}
    \item Denote all the fields in the bulk theory by $\psi_i$.
    \item The bulk action $S_c$ has two components, (1) a bulk Lagrangian density $\mathcal{L}$ and (2) possibly a boundary term with a boundary density $\mathcal{L}_{\text{bdy}}$.
    \begin{align}
        S_c = S_{\text{bdy}} + \int{\text{d}^{d+1} y \sqrt{g} \mathcal{L}}.
    \end{align}
    \item The functional derivative therefore becomes
    \begin{align}
        \frac{\delta S_c(\psi_{(0)})}{\delta \psi_{(0)}(x)} |_{\psi_(0)=0} = &\int{\text{d}^{d+1}y \sqrt{g} ( \frac{\partial \mathcal{L}(y)}{\partial \psi_i(y)} \frac{\partial \psi_i(y)}{\partial \psi_{(0)}(x)}} \nonumber \\ &+ \frac{\partial \mathcal{L}(y)}{\partial(\nabla_a \psi_i(y))} \frac{\partial(\nabla_a \psi_i(y))}{\partial \psi_{(0)}(x)} ) \nonumber \\ &+ \frac{\delta S_{\text{bdy}}}{\delta \psi_{0}(x)}|_{\psi_{(0)}=0},
    \end{align}
    where $i$ goes over all fields and summation is implied as in the Einstein summation convention, with a bulk Lagrangian assumed to only depend on fields and their first derivatives.
    \item Now integrate by parts such that
\begin{center}
\begin{align}
        \frac{\delta S_c(\psi_{(0)})}{\delta \psi_{(0)}(x)}|_{\psi_{(0)}=0} &= \int \text{d}^{d+1} y \sqrt{g}  ( \frac{\partial \mathcal{L}(y)}{\partial \psi_i(y)} \nonumber \\ &- \nabla_a \frac{\partial \mathcal{L}(y)}{\partial(\nabla_a \psi_i(y))})  \frac{\partial \psi_i(y)}{\partial \psi_{(0)}(x)} \nonumber \\ &+ \int_{\partial \text{AdS}} \text{d}^d y \sqrt{g_{(0)}} n_a \frac{\partial \mathcal{L}(y)}{\partial(\nabla_a \psi_i(y))} \frac{\partial \psi_i(y)}{\partial \psi_{(0)}(x)} \nonumber \\  &+ \frac{\delta S_{\text{bdy}}}{\delta \psi_{(0)}(x)}|_{\psi_{(0)} = 0},
\end{align}
\end{center}
where $n_a$ is an outward normal to the boundary of AdS. The first integral vanishes because the fields obey the Euler-Lagrange equation, and the CFT expectation value can therefore be read off from the boundary behavior of the bulk fields through this expanded relation, given Equation $~\ref{eqn:expectation}$.
\end{enumerate}
\ \ \ \ A CFT expectation value can then, via this process, be obtained from each of the bulk fields once the boundary behavior of the on-shell bulk fields are known. The relation
\begin{align}
    \braket{\mathcal{O}}_{\text{CFT}} &= \int_{\partial \text{AdS}} \text{d}^d y \sqrt{g_{(0)}} n_a \frac{\partial \mathcal{L}(y)}{\partial(\nabla_a \psi_i(y))} \frac{\partial \psi_i(y)}{\partial \psi_{(0)}} \nonumber \\ &+ \frac{\delta S_{\text{bdy}}}{\delta \psi_{(0)}(x)}|_{\psi_{(0)}=0}
    \label{eqn:CFT_expec}
\end{align}
will be the basis of much of the upcoming mathematical framework, as later demonstrated.

In said framework, however, a strongly coupled boundary theory will be considered. Conformal field theories are characterized by not having any particular length scale and the physics of critical points, which include thermodynamic and quantum phase transitions, often has this property. The physics near a critical point can be expected  as similar to the critical system and finding the critical behavior is then of interest.

\section{\label{sec:level1}Mathematical Approach Overview}
The AdS/CFT correspondence will be used to model a high-$T_c$ superconductor, both below and above $T_c$. This will first be demonstrated through the simplest possible Lagrangian, with a determined frequency-dependent conductivity. The Lagrangian is then extended with a higher-order term and outlying behavior will be identified. Finally, the Drude model will be used to further describe and explain outlying trends in conductivity.

\section{\label{sec:level1}Two-Dimensional Condensed Matter Systems}
We are to model a superconductor with a high $T_c$. While conventional superconductors are well-described by the BCS (Bardeen–Cooper–Schrieffer) theory, in which electrons, photons, and phonons are the degrees of freedom of interest. The importance of phonon interactions was understood from the isotope effect, the mass of the atoms in the lattice changing the superconductivity behavior. The isotope effect is much weaker, however in high-temperature superconductors and the phonons are thus not believed to be important for high temperature superconductivity\cite{leggett2006we}. The important degrees of freedoms are the electrons and photons. These electrons are, as in BCS theory, expected to form Cooper pairs\cite{leggett2006we}, pairs of electrons of opposite spin but otherwise effectively in the same state becoming spin 0 particles.The methodology presented here is adapted from the holographic methods review by P. Säterskog, who computed the numerics shown throughout this section using a Prince-Dormand method\cite{Saterskog13}, to which end our superconductor model will thus contain two fields: a spin 1 field $A_a$ for the photons and a spin 0 field $\psi$ for the Cooper-pairs.
\subsection{\label{sec:level1}Symmetry assumptions}
\ \ \ \ The bulk theory should present the same symmetries as the boundary theory. A U(1) gauge symmetry of the complex $\psi$ field will therefore be imposed. Lorentz invariance will also be employed for both theories, even though relativistic phenomena are largely negligible for superconductivity. 
\subsection{\label{sec:level1}A bulk Lagrangian}
\ \ \ \ There are many different possible bulk Lagrangians for the fields $A_a$ and $\psi$ and the metric $g_{ab}$. A Lagrangian previously successfully used to model two-dimensional electron condensates,
\begin{align}
    \mathcal{L} = \frac{1}{2 \kappa} (R - 2 \Lambda) - \frac{1}{4} F_{ab} F^{ab} - m^2 |\psi|^2 - |D_a \psi|^2,
    \label{eqn:first_lagrangian}
\end{align}
having been obtained through Wilsonian naturalness\cite{hartnoll9, horowitz}. Of Equation $~\ref{eqn:first_lagrangian}$, the first term is an Einstein-Hilbert term with cosmological constant $\Lambda$, with a negative constant giving the required asymptotically AdS space, the Ricci scalar curvature $R$, and coupling constant between metric and other fields $\kappa$; the second term is an ordinary Maxwell term where electromagnetic tensor $F_{ab}$ is the exterior derivative of the electromagnetic field tensor $F_{ab} = \partial_a A_b - \partial_b A_a$; and the third and fourth terms are the kinetic and mass terms for the scalar field, respectively, with covariant derivative $\nabla_a$ and gauge covariant derivative $D_a = \nabla_a - iq A_a$. The action of this Lagrangian is
\begin{align}
    S = \int{\text{d}^{d+1} x \sqrt{g} \mathcal{L} + S_{\text{boundary}}},
\end{align}
with $g$ as the absolute value of the determinant of the metric tensor $g = |\text{det}( g_{ab})|$. $S_{\text{boundary}}$ is a boundary term needed to cancel divergences when integrating the action towards the boundary. It does not significantly affect equations of motion but is needed to get normalizeable modes. 

In accordance with the previous symmetry assumptions, the described gauge coupling makes the Lagrangian invariant under the U(1) gauge transformation
\begin{align}
    \psi &\rightarrow e^{i \theta(x)} \psi \\ A_a &\rightarrow A_a + \frac{1}{q} \nabla_a \theta(x).
\end{align}
This Lagrangian is also manifestly Lorentz invariant, imposing the Lorentz invariance of the boundary theory.
\subsection{\label{sec:level1}Equations of motion}
The bulk equations of motion are obtained by varying the bulk Lagrangian with respect to all fields, a procedure possible through Noether's theorem. This can be done with the Euler-Lagrange equation since the action does not contain any higher derivatives. The Euler-Lagrange equation for a scalar field $\chi$ states that
\begin{align}
    \nabla_a (\frac{\partial \mathcal{L}}{\partial(\nabla_a \chi)}) - \frac{\partial \mathcal{L}}{\partial \chi} = 0.
\end{align}
We first vary $\psi$, giving
\begin{align}
    (m^2 - \nabla^2 + q^2 A^2 + iq(\nabla_a A^a)) \psi = 0.
\end{align}
Varying $A_a$ gives the equations of motion
\begin{align}
    -\nabla_a F^{ab} + 2q^2 |\psi|^2 A^b + iq(\overline{\psi} \nabla^b \psi - \psi \nabla^b \overline{\psi}) = 0.
\end{align}
A real $\psi$ simplifies calculations and can be obtained since the gauge invariance lets us relate any configuration to a real one through a gauge transformation. The Lorentz gauge 
\begin{align}
    \nabla_a A^a = 0,
    \label{eqn:Lorentz_gauge_condition}
\end{align}
removes the last term in the parentheses of the equation of motion for $\psi$. The equation of motion for $\psi$
does not mix the real and imaginary parts after this choice and $\psi$ can then be taken to be real since a global shift of phase does not affect $A_a$. The gauge is still not completely fixed, a gauge transformation $\theta(x)$ such that $\nabla_a \nabla^a \theta(x) = 0$ can still be done without violating the gauge condition of Equation $~\ref{eqn:Lorentz_gauge_condition}$. After choosing the Lorentz gauge and a real $\psi$, the equations of motion are
\begin{align}
    \label{eqn:motion}
    \begin{split}
        (m^2 - \nabla^2 + q^2 A^2) \psi = 0, \\ -\nabla_a F^{ab} + 2q^2 \psi^2 A^b = 0.
    \end{split}
\end{align}
\subsection{\label{sec:level1}Lagrangian Parameters}

As seen, there are four free parameters in the bulk Lagrangian: $\kappa$, $\Lambda$, $m^2$, and $q$. These must be investigated to find values that give us the boundary theory we are interested in. Some of these parameters may be redundant since we can make different symmetry transformations of fields and coordinates. The physics of the bulk are treated in the classical limit and the Lagrangian can thus be changed as long as the equations of motion for $\psi$ and $A_a$ are left unchanged.
\subsubsection{\label{sec:level1} On the $\kappa$ Parameter}
This Einstein-Hilbert term makes the theory employed here gravitational. $\kappa$ is proportional to Newton's gravitational constant, with a small $\kappa$ giving the probe limit where the metric equations of motion can be solved independently of the other fields, although it is not guaranteed here that boundary theories of interest are dual to bulk theories. This can be understood by varying the Lagrangian with respect to the metric; the Einstein-Hilbert part gives a term inversely proportional to $\kappa$ and the rest of the Lagrangian gives the stress-energy tensor independently of $\kappa$.

This greatly simplifies calculations and will therefore be used throughout this investigation. 

\subsubsection{\label{sec:level1} On the $\Lambda$ Parameter}
The scale invariance of the system allows an arbitrary $\Lambda$, as two systems with different cosmological constants can be equated through a rescaling. 
It will later be shown that the cosmological constant and length scale $L$ can be related via
\begin{align}
    L = \sqrt{-\frac{3}{\Lambda}},
\end{align}
which also allows for other parameters to be related and scale invariance can thereby not be used to choose these parameters freely. 

\subsubsection{\label{sec:level1} On the $q$ Parameter}
The gauge coupling strength $q$ is also the charge of the scalar field, hence, the letter used. Taking conjugates $\tilde{\psi} = q \psi$ and $\tilde{A}_a = q A_a$ as fields will give a Lagrangian identical to our previous Equation $~\ref{eqn:first_lagrangian}$, but divided by $q^2$ except for the term originally containing $q^2$, which is divided by $q^4$. Scaling the Lagrangian by a constant does not affect the equations of motion, so the system can be solved for any value of $q$. Other solutions can be obtained by rescaling the fields.
\subsubsection{\label{sec:level1} On the $m^2$ Parameter}
The values of $m$, the mass of the scalar field in the bulk, that are suitable will later be investigated when solving equations of motion in the bulk.
\section{\label{sec:level1}Solutions of the Classical Bulk Theory}
The Lagrangian describes a general system, so there exist many solutions to the described equations of motion. We wish to investigate two properties of a superconductor, the development of a condensate at low temperatures and the conductivity at different frequencies. We are interested in a superconductor subject to spatially uniform conditions, the applied electric field uniform and the chemical potential uniform. The atomic lattice and its imperfections are thus not accounted for, but interesting superconductivity behavior can be obtained anyway\cite{hartnoll9}. It is thus enough to examine a system with translational symmetry in the $x$ and $y$ directions. A rotationally invariant superconductor will further be studied. The system is subject to conditions constant in time, such as there being no time-dependent chemical potential. This allows for the assumption of time independence when solving the non-linear field equations.

The conductivity is the linear electrical current response to an applied transverse electrical field. We apply this in the $x$ direction due to the rotational symmetry and let the applied field have a harmonic time dependence $\exp(i t \omega)$ so we can get the response function in the frequency domain. The linear response is sought so the applied field should be infinitesimal. The applied field breaks the rotational and time symmetries but since it is infinitesimal and we are not interested in the effect it has on another fields, it can be neglected in calculations. The applied field is later added with the other fields as a background solution.

The electrical field in the $x$-direction is $E_x = F_{xt} = \partial_x A_t - \partial_t A_x$. Translational symmetry gives $E_x = \partial_t A_x$. These limitations allow for the following definitions:
\begin{align}
    \text{d}s^2 &= g_{tt}(z) \text{d}t^2 + g_{xx}(z)(\text{d}x^2 + \text{d}y^2) + g_{zz}(z) \text{d}z^2, \\ \psi &= \psi(z),
\end{align}
and
\begin{align}
    A_a = (\phi(z), A_x(z) \exp(it \omega),0,A_z(z)),
\end{align}
where $\phi(z)$ is infinitesimal. The gauge condition requires
\begin{align}
    \nabla_a A^a = \partial_a A^a + \Gamma^a_{ba} A^b = 0,
\end{align}
which gives a homogeneous first-order linear ordinary differential equation for $A_z$ because the contracted Christoffel symbol only has a $z$ component. The remaining gauge symmetry lets us add a function to $A_z$ and can be used to set $A_z(z) = 0$ for a specific $z$. The above differential equation then requires $A_z(z)$ to be identically 0 for all $z$. We will hereafter work with $A_z(z) = 0$.

The explicit $z$- and $t$-dependence of these functions will hereafter be omitted.

\subsection{\label{sec:level1}The Schwarzchild metric}

The path integral for the bulk partition function is approximated in a semi-classical approximation where we need the saddle point of the action. We first wish to find the metric saddle point of the periodic imaginary time path integral. The bulk equation of motion for the metric $g_{ab}$ is the Einstein equation
\begin{align}
    R_{ab} - \frac{1}{2} g_{ab} R + g_{ab} \Lambda = \kappa T_{ab},
    \label{eqn:Einstein_equation}
\end{align}
where $R_{ab}$ is the Ricci curvature tensor and $T_{ab}$ is the stress-energy tensor. The probe limit was assumed, so we may assume a small $\kappa$ and neglect the right-hand side of Equation $~\ref{eqn:Einstein_equation}$. We desire a translation-invariant solution in the $t$, $x$, and $y$ directions that is asymptotically AdS. The solution is known to be a black hole\cite{McGreevy:2009xe}, the Schwarzchild metric in AdS space. The metric has the following form in a particular choice of coordinates where the radial coordinate $z$ is 0 at the boundary and $z_h$ at the horizon,
\begin{align}
    g_{ab} \text{d}x^a \text{d}x^b = \frac{L^2}{z^2}(\frac{\text{d}z^2}{f(z)} - f(z) \text{d}t^2 + \text{d}x^2 + \text{d}y^2),
    \label{eqn:metric}
\end{align}
where $f(z) = 1 - z^3 z_h^{-3}$. This function of $z$ approaches $1$ at the boundary and the space is asymptotically AdS. There is a horizon at $z = z_h$ where $f(z_h) = 0$. The space behind the horizon can not affect the physics of the boundary and can therefore be neglected in these calculations. This solution is periodic in imaginary time. Consider the near-horizon metric where
\begin{align}
    f(z) = f(z_h) - (z_h - z)f'(z_h) + \mathcal{O}((z_h - z)^2) \approx 3(1 - zz_h^{-1})
\end{align}.
With a change of variables of $\rho^2 = (4L^2)/3(1 - zz_h^{-1})$, we have $f(z) \approx 9 \rho^2/4L^2$ and $\rho^2 \text{d}\rho^2 = \text{d}z^2 z_h^{-2} \cdot 4L^4/9$. The near-horizon metric is then
\begin{align}
    g_{ab} \text{d}x^a \text{d}x^b = \frac{L^2}{z_h^2}(\frac{\rho^2 \text{d}\rho^2}{z_h^{-2} \cdot \frac{4L^4}{9} \rho^2 \frac{9}{4L^2}} - \rho^2 \frac{9}{4L^2} \text{d}t^2 + \text{d}x^2 + \text{d}y^2).
\end{align}
We now extend to the imaginary time $\tau = it$, where
\begin{align}
    g_{ab} \text{d}x^a \text{d}x^b = \text{d}\rho^2 + \rho^2(\frac{3}{2 z_h} \text{d} \tau)^2 + \frac{L^2}{z_h^2}(\text{d}x^2 + \text{d}y^2).
\end{align}
The near-horizon metric is then that of a Euclidean plane in polar coordinates, and there is thus a deficit angle unless $(3/2z_h)\tau$ has a periodicity of $2 \pi$. The imaginary time has periodicity $\beta$, so we must have
\begin{align}
    \frac{3}{2z_h} = \frac{2\pi}{\beta}.
\end{align}
This gives the relationship between $z_h$ and the temperature
\begin{align}
    T = \frac{3}{4 \pi z_h}.
    \label{eqn:temp}
\end{align}
This expression for the temperature agrees with the Beckenstein-Hawking temperature of a black hole.

We have assumed $\kappa T_{ab} = 0$ in finding the metric. The backreaction, $\delta g_{ab}$, from the non-zero fields will be of order $\kappa T_{ab}$, as according to Equation $~\ref{eqn:Einstein_equation}$. The Einstein equation is obtained by varying the Lagrangian with respect to $g_{ab}$ and finding the saddle point so $\delta S \propto \kappa^{-1} \delta g_{ab}^2$ for the variation around the saddle point. We thus have that $\delta S \propto \kappa T^2_{ab}$ and the backreaction can be safely neglected when calculating the action from different configurations.

The background metric can now be used instead of solving the equations of motion for the metric together with the fields. The gravitational part of the Lagrangian must be kept when calculating the value of the total action, which is dominated by the gravitational part.

The horizon $z_h$ and the curvature length $L$ set length scales in the metric. Length units in the numerical solution can be chosen such that $z_h = 1$. This means that for different temperatures, we have different units, since $z_h$ is related to the temperature. We will have to convert between these units when comparing results from different temperatures.

\subsection{\label{sec:level1}Field equations of motion}

The equations of motion for $\psi(z)$, $\phi(z)$, and $A_x(z)$ can now be obtained. With the past definitions for $\text{d} s^2$, $\psi$, and $A_a$, the equations of motion of Equation $~\ref{eqm1}$ through $~\ref{eqm3}$, and the metric of Equation $~\ref{eqn:metric}$, we have

\begin{align}\Big(q^2z^2\phi^2-L^2m^2f+zf(zf^\prime-2f)\partial_z+z^2f^2\partial_z\partial_z\Big)\psi=0\label{eqm1}\end{align}
 \begin{align}\Big(-2q^2\psi^2L^2+z^2f\partial_z\partial_z\Big)\phi=0\label{eqm2}\end{align}
 \begin{align}\Big(-2q^2\psi^2L^2f+z^2\omega^2+z^2ff^\prime\partial_z+z^2f^2\partial_z\partial_z\Big)A_x=0\label{eqm3}.
\end{align}
A Frobenius expansion \cite{teschl2012ordinary} of these equations can be done at the boundary, $z=0$. The leading behavior of the functions is
\begin{empheq}[left=\empheqlbrace]{align}
 &\psi=\psi_{(0)}\left(\frac{z}{L}\right)^{\Delta_\psi}\\
 &\phi=\phi_{(0)}\left(\frac{z}{L}\right)^{\Delta_\phi}\\
 &A_x=A_{x(0)}\left(\frac{z}{L}\right)^{\Delta_{A_x}}
\end{empheq}
where $\Delta_\psi$, $\Delta_\phi$ and $\Delta_{A_x}$ are constants that are to be determined. This is a slight assumption since not all functions have this type of leading behavior\footnote{The function $\log$ for example does not allow an expansion like this, we have assumed the function does not have an essential singularity at $z=0$.}.
 Entering this in the equations of motion yields
 \begin{align}
  q^2z^2\phi_{(0)}^2s^{2\Delta_\phi} &- \nonumber \\ L^2m^2f &+ \nonumber \\  f(zf^\prime-2f)\Delta_\psi &+ \nonumber \\ f^2\Delta_\psi(\Delta_\psi-1) &=0,\label{ind1}\end{align}
  \begin{align}-2q^2\psi_{(0)}^2s^{2\Delta_\psi}L^2+f\Delta_\phi(\Delta_\phi-1)=0,\label{ind2}\end{align}
and 
  \begin{align}-2q^2\psi_{(0)}^2s^{2\Delta_\psi}L^2f &+ \nonumber \\ z^2\omega^2 &+ \nonumber \\ zff^\prime\Delta_{A_x} &+ \nonumber \\ f^2\Delta_{A_x}(\Delta_{A_x}-1)&=0.
 \end{align}
where $s=zL^{-1}$. This immediately gives $\Delta_\psi\geq0$ and $1+\Delta_\phi\geq0$ since the first terms otherwise diverge at the horizon where the other terms are finite. First consider the case of strict inequalities. The leading order behavior is then
 \begin{empheq}[left=\empheqlbrace]{align}
  &-L^2m^2-2\Delta_\psi+\Delta_\psi(\Delta_\psi-1) =0\\
  &\Delta_\phi(\Delta_\phi-1)=0\\
  &\Delta_{A_x}(\Delta_{A_x}-1)=0.
 \end{empheq}
with solutions
 \begin{empheq}[left=\empheqlbrace]{align}
  &\Delta_\psi =\frac{3}{2}\pm\sqrt{\frac{9}{4}+L^2m^2}\label{indicialSo1}\\
  &\Delta_\phi=0,1\\
  &\Delta_{A_x}=0,1\label{indicialSol3}.
 \end{empheq}
Observe that each of these three exponents have two solutions, each \emph{independent} of one another.
Now assume $\Delta_\psi=0$. Equation \ref{ind2} gives
\begin{equation}
 -2q^2\psi_{(0)}^2L^2+\Delta_\phi(\Delta_\phi-1)=0
\end{equation}
while Equation \ref{ind1} gives $\Delta_\phi=-1$. We then have
 \begin{empheq}[left=\empheqlbrace]{align}
  &q^2\phi_{(0)}^2z_h^{2} =L^2m^2\label{spurious1}\\
  &q^2\psi_{(0)}^2L^2=1\\
  &\Delta_{A_x}(\Delta_{A_x}-1)=2.
 \end{empheq}
with solutions $\Delta_{A_x}=-1,2$.
First assuming $\Delta_\phi=-1$ yields the same result. There are however no solutions to Equation \ref{spurious1} for the negative $m^2$ we later will consider and infinities are encountered when calculating the action for these solutions so they will not be considered.
All useful solutions are thus given by Equation \ref{indicialSo1} to Equation \ref{indicialSol3}.
\subsection{Field behavior at horizon\label{s:hb}}
The same kind of Frobenious expansion can be made at the horizon but there are some simplifying conditions. The time component of the metric vanishes at the horizon, $f(z_h)=0$. This means that $A_t(z_h)$ must be zero because a finite $A_t(z_h)$ would give a finite Wilson loop around the periodic imaginary time circle whose length in time is 0. A Wilson loop is contrary to $A_t$ a physical quantity ($A_t$ is gauge-dependent). This gives a singular gauge connection which is unphysical \cite{hartnoll8}. We thus have $A_t(z_h)=0$. Expand the fields as 
\begin{empheq}[left=\empheqlbrace]{align}
 &\psi=\psi_{(h)}s^{\Delta^{(h)}_\psi}\\
 &\phi=\phi_{(h)}s^{\Delta^{(h)}_\phi}\\
 &A_x=A_{x(h)}s^{\Delta^{(h)}_{A_x}}
\end{empheq}
where $s$ now is $(1-z/z_h)$ and $\Delta^{(h)}_\phi>0$\footnote{The notation $\Delta^{(h)}$ is used to signify that these exponents describe the \emph{horizon} behavior, $\Delta$ was earlier used for the conformal boundary behavior.}. The function $f$ can be expanded as $f=3s-3s^2+s^3$.
% \begin{empheq}[left=\empheqlbrace]{align}
%  &q^2z^2\phi_{(1)}^2s^{2\Delta^{(h)}_\phi}-L^2m^2f-zf(zf^\prime-2f)s^{-1}\Delta^{(h)}_\psi+z^2f^2s^{-2}\Delta^{(h)}_\psi(\Delta^{(h)}_\psi-1)=0\\
%  &-2q^2\psi_{(1)}^2s^{2\Delta^{(h)}_\psi}L^2+z^2fs^{-2}\Delta^{(h)}_\phi(\Delta^{(h)}_\phi-1)=0\\
%  &-2q^2\psi_{(1)}^2s^{2\Delta^{(h)}_\psi}L^2f+z^2\omega^2-z^2ff^\prime s^{-1}\Delta^{(h)}_{A_x}+z^2f^2s^{-2}\Delta^{(h)}_{A_x}(\Delta^{(h)}_{A_x}-1)=0
% \end{empheq}
Inserting these leading terms in the equations of motion gives
\begin{align}
 q^2z^2\phi_{(h)}^2s^{2\Delta^{(h)}_\phi} &+ \nonumber \\ 9z_h^2\Delta^{(h)}_\psi &+ \nonumber \\ z_h^2 9\Delta^{(h)}_\psi(\Delta^{(h)}_\psi-1)&=0,\end{align}
 
 \begin{align}
 -2q^2\psi_{(h)}^2s^{2\Delta^{(h)}_\psi}L^2 &+ \nonumber \\ z_h^2 3s^{-1}\Delta^{(h)}_\phi(\Delta^{(h)}_\phi-1)&=0,\end{align}
 and
 \begin{align}
 -6q^2\psi_{(h)}^2s^{2\Delta^{(h)}_\psi+1}L^2 &+ \nonumber \\ z_h^2\omega^2 &+ \nonumber \\  9\Delta^{(h)}_{A_x} &+ \nonumber \\ 9\Delta^{(h)}_{A_x}(\Delta^{(h)}_{A_x}-1)&=0. \end{align}
Solving for the leading terms of these equations and using $\Delta^{(h)}_\phi>0$ gives
% \begin{empheq}[left=\empheqlbrace]{align}
%  &\Delta^{(h)}_\psi=0\\
%  &\Delta^{(h)}_\phi(\Delta^{(h)}_\phi-1)=0\\
%  &\omega^2+  9\Delta^{(h)}_{A_x}^2=0.
% \end{empheq}
% This has solutions
\begin{empheq}[left=\empheqlbrace]{align}
 &\Delta^{(h)}_\psi=0\\
 &\Delta^{(h)}_\phi=1\\
 &\Delta^{(h)}_{A_x}=\pm\frac{\i\omega z_h}{3}\label{inout}.
\end{empheq}
The two possible $\Delta^{(h)}_{A_x}$ represent solutions going into or coming out of the horizon. Close to the horizon is $A_x(z,t)$ given by
\begin{equation}
 A_x(z,t)=s^{\pm\frac{\i\omega z_h}{3}}\exp(\i\omega t)=\exp\big(\i\omega(t\pm\frac{z_h\log s}{3})\big).
\end{equation}
The phase is constant for $s=\exp(\mp3t/z_h)$ so the plus sign in Equation \ref{inout} gives the ingoing solution.

\subsubsection{\label{sec:level1}Boundary conditions}
The equations of motion, Equations $~\ref{eqm1}$, $~\ref{eqm2}$, and $~\ref{eqm3}$, can be integrated numerically. Just one leading horizon behavior is allowed for $\psi$ and $\phi$, so only two horizon conditions are needed for them, $\psi(z_h)$ and $\phi'(z_h)$. The derivative $\psi'(h)$ needed for starting a numerical integration from the horizon can be obtained directly from the equations of motion as $z \rightarrow 0$:
\begin{align}
    \psi'(z_h) = -\frac{L^2 m^2}{3z_h}.
\end{align}
A two-parameter family of solutions to the equations of motion can then be obtained for $\psi$ and $\phi$. These solutions give the boundary values of the fields which describe the background fields of the field theory. The two horizon parameters must be chosen to obtain the desired background fields.

The field theory operator $\mathcal{O}$ corresponding to the background field $\psi$ is expected to spontaneously attain a non-zero expectation value breaking the U(1) symmetry, so we therefore require the source $\psi_{(0)} = 0$. The time component of the electromagnetic potential $A_a$ corresponds to the electric potential in the Lorentz gauge. The electrical potential gives the energy per charge needed to add a charge to the system and the chemical potential for the electrons $\mu$ can thus be expressed as $\mu = q \psi_{(0)}$. 

Just one of the two horizon behaviors of the Maxwell perturbation $A_x$ is wanted. We want a casual response from the perturbation of the background field. This corresponds to the solution going into the horizon as time passes\cite{hartnoll8}. We thus choose the ingoing horizon behavior. The equation for $A_x$ is linear and we are only interested in the linear response at the conformal boundary, so the horizon amplitude of the ingoing solution can be chosen arbitrarily.

The horizon parameters $\psi(z_h)$ and $\phi'(z_h)$ can now be varied to find solutions to the two boundary conditions $\psi_{(0)} = 0$ and $\mu = q \psi_{(0)}$.

There is a trivial analytical solution of the equations of motion with the above boundary conditions.
\begin{equation}
\begin{cases} 
 &\psi(z)=0\\
 &\phi(z)=\mu(1-z/z_h)\\
 &A_x(z)=
\left[
\exp\left( - \sqrt{3}\tan^{-1} \frac{z_h+2z}{z_h\sqrt{3}} \right)
\frac{z_h-z}{\sqrt{z^2+zz_h+z_h^2}}
\right]^{\frac{\i\omega z_h}{3} }   \label{trivial}
\end{cases}
\end{equation}

The field $\psi$ is here identically zero and there is no spontaneous symmetry breaking. This solution thus corresponds to the physics above the critical temperature, $T_c$, of the superconductor. We will now make a choice of $m$ to be able to numerically investigate solutions with $\psi \neq 0$.

\subsubsection{\label{sec:level1}Choice of scalar mass $m$}
The mass squared of a scalar field in flat space must be non-negative for stability. This is not the case in a space with negative curvature, however. The Breitenlohner-Freedman (BF) bound is a lower stability bound on $m^2$ of a massive scalar field in AdS space with metric given by Equation $~\ref{eqn:metric}$. It requires that\cite{Kleban:2004bv}
\begin{align}
    L^2 m^2 \geq -\frac{d^2}{4}.
\end{align}
The scalar field $\psi$ should obey this bound far away from the black hole for normalizable modes. We would however like a spontaneous symmetry breaking of $\psi$ near the black hole corresponding to the electron condensate\cite{Gubser:2008px}. This can happen because the coupling of $\psi$ to the perturbation $A_a$
gives $\psi$ an effective mass that might break the BF bound near the black hole. The effective mass is given by
\begin{align}
    m_{\text{eff}}^2 = m^2 + A_a A^a = m^2 - \frac{z^2}{L^2(1 - z^d z_h^{-d})} \phi^2.
    \label{eqn:eff_mass}
\end{align}
This can for large enough values of $\phi$ break the BF bound. Let us consider the trivial, uncondensed solution of Equation $~\ref{trivial}$. When does this give an effective mass breaking the BF bound and possibly enabling an additional condensed solution? The location of the effective mass minimum, $z_0$, can be found by differentiating the above Equation $~\ref{eqn:eff_mass}$ by $z$ and using the uncondensed solution of Equation $~\ref{trivial}$,
\begin{align}
    \frac{z_0}{z_h} = \frac{1}{3}(\sqrt[3]{37 +9 \sqrt{17}} - \frac{2}{\sqrt[3]{37 + 9 \sqrt{17}}} - 2).
\end{align}
The effective mass breaks the BF bound at $z_0$ when
\begin{align}
    \frac{\mu}{T} > \frac{2 \pi}{\sqrt{3}} \sqrt{\frac{4L^2 m^2 + 9}{8 - \sqrt[3]{142 - 34 \sqrt{17}} - \sqrt[3]{142 + 34 \sqrt{17}}}},
\end{align}
where Equation $~\ref{eqn:temp}$ has been used. We will choose $m^2L^2 = -2$, which does not break the BF bound, but is relatively close. It gives integer scalings for the scalar field at the conformal boundary, which is convenient.

\subsection{\label{sec:level1}Expectation values of field theory operators}

Expectation values of field theory operators can now be calculated using the solutions of the equations of motion of Equation $~\ref{eqm1}$, $~\ref{eqm2}$, and $~\ref{eqm3}$, and the previously discussed CFT expectation value of Equation $~\ref{eqn:CFT_expec}$. We will later see, however, that not only is the leading behavior of the fields needed to calculate the expectation values, but also the first subleading behavior. We therefore expand the fields as
\begin{empheq}[left=\empheqlbrace]{align}
 &\psi=\psi_{(0)} \frac{z}{L} + \psi_{(1)} (\frac{z}{L})^2\\
 &\phi = \phi_{(0)} + \phi_{(1)} \frac{z}{L}\\
 &A_x = A_{x(0)} + A_{x(1)} \frac{z}{L}
\end{empheq}
and obtain $\psi_{(i)}$, $\phi_{(i)}$, and $A_{x(i)}$ from the numerical solution. For this the boundary terms of the action are required. The boundary term needed for the scalar field is calculated as
\begin{align}
    S_{\text{bdy}} = -\int_{z=\epsilon}{\text{d}^d x L^{-1} \psi^2 \sqrt{g_{(0)}}}.
\end{align}
We may insert this in Equation $~\ref{eqn:CFT_expec}$, so 
\begin{align}
    \braket{\mathcal{O}}_{\text{CFT}} &= -\int_{\partial \text{AdS}}{\text{d}^d y \sqrt{g_{(0)}} (n_a \frac{\partial \mathcal{L}(y)}{\partial(\nabla_a \psi(y))} - 2L^{-1} \psi) \frac{\delta \psi(y)}{\delta \psi_{(0)}(x)}} \nonumber \\ &= \frac{L^3}{z^3} (\frac{z}{L} 2 \nabla_z \psi(y) - 2L^{-1} \psi) \frac{z}{L} \nonumber \\ &= \frac{L^2}{z^2} (\frac{z}{L} 2(\psi_{(0)} \frac{1}{L} + \psi_{(1)} \frac{2z}{L^2}) - 2L^{-1}(\psi_{(0)} \frac{z}{L} + \psi_{(1)} (\frac{z}{L})^2) \nonumber \\ &= \frac{2 \psi_{(1)}}{L}.
\end{align}
This simple relationship gives us the expectation value of the scalar operator. 

We are now in a position to numerically solve the bulk theory and obtain the expectation values. We do this by sweeping over different horizon values and for each value finding all solutions to the boundary condition $\psi_{(0)} = 0$. This yields many different solutions $\rho/T$ at the boundary. Scale invariance lets us interpret this as systems of constant $\rho$ but at different temperatures $
T$. We then get a variation in the chemical potential $\mu$. The chemical potential of the trivial solution shown there is calculated through
\begin{align}
    \rho = \frac{\mu}{z_h} = \mu T \frac{4 \pi}{3}.
    \label{eqn:chem_trivial}
\end{align}
Alternatively, one can let $\mu$ be constant while varying the temperature and get a variation in $\rho$. 

\begin{figure}
    \centering
    \includegraphics[width=0.50\textwidth]{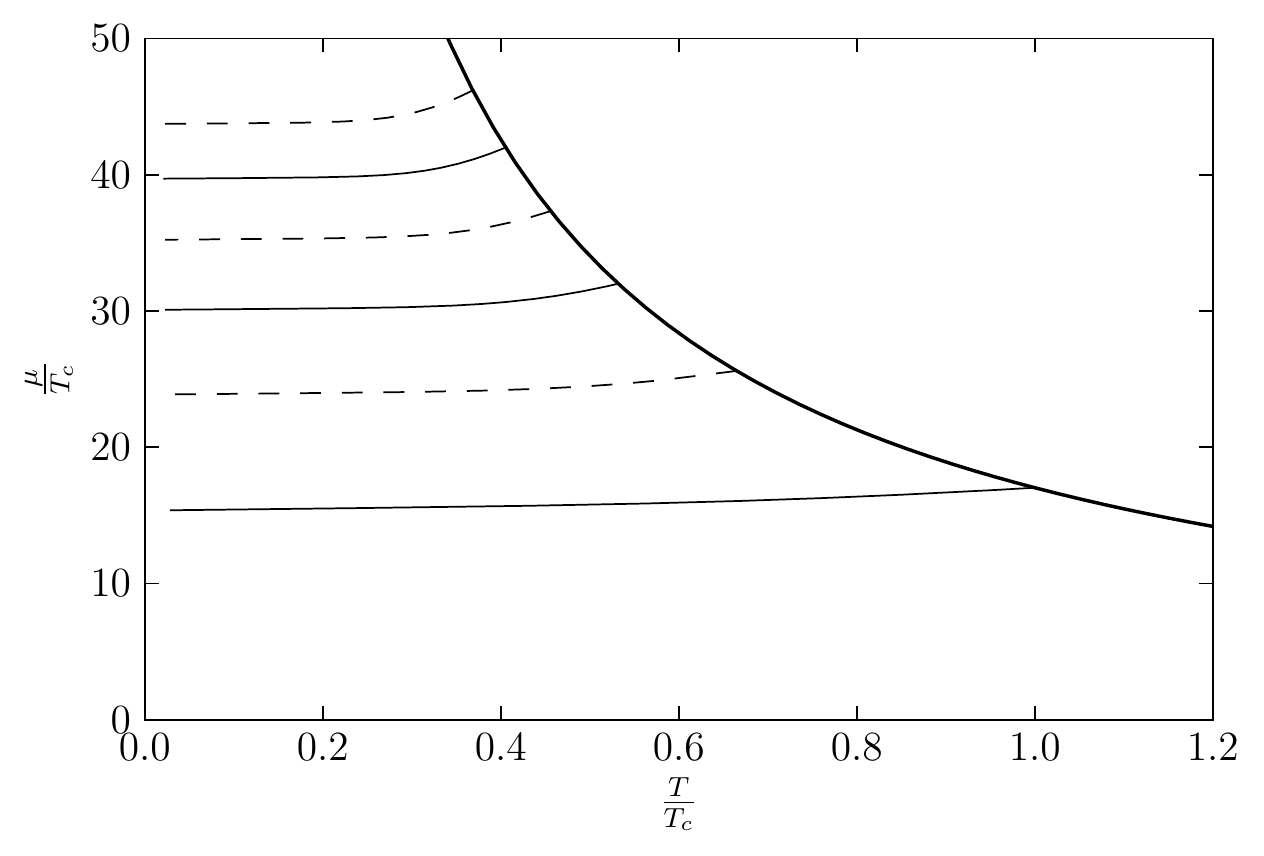}
    \caption{Chemical potential needed for constant $\rho$ at different $T$. The multiple curves correspond to multiple solutions at the same temperature. The dashed lines have different signs of the expectation value of $\mathcal{O}$ and the horizon boundary condition $\psi(z_h)$. Further solutions (here omitted) are obtained for lower temperature following the trend shown here.}
    \label{fig:mu}
\end{figure}

\begin{figure}
    \centering
    \includegraphics[width=0.50\textwidth]{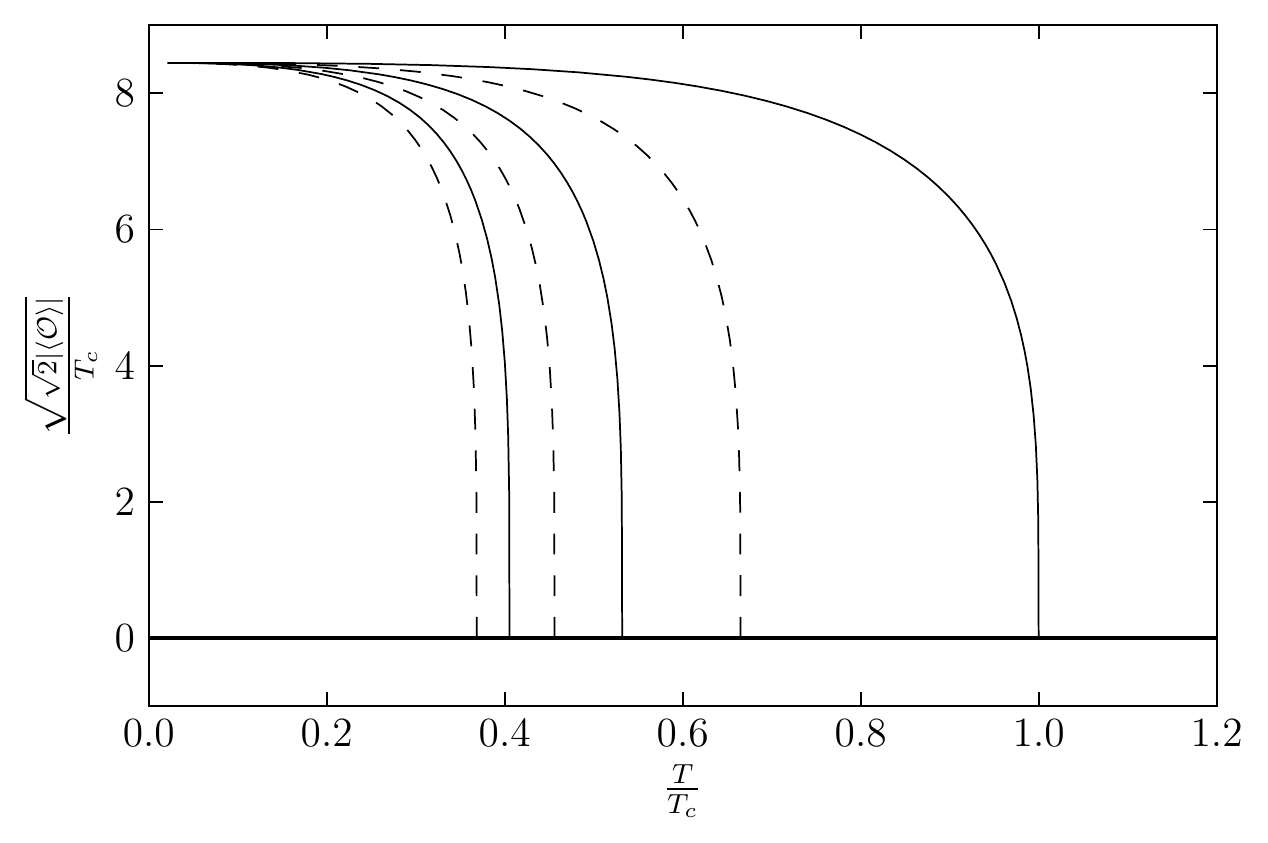}
    \caption{Expectation value of CFT operator $\mathcal{O}$ at different $T$ and constant $\mu$. The multiple curves correspond to multiple solutions at the same temperature. The dashed lines have different signs of the expectation value of $\mathcal{O}$ and the horizon boundary condition $\psi(z_h)$. Further solutions (here omitted) are obtained for lower temperature following the trend shown here.}
    \label{f:O}
\end{figure}

\subsection{\label{sec:level1}Free energy and electrical conductivity}

The free energy, $A = -T \log{Z}$, is the same for the bulk and the boundary theory since their partition functions are the same. This can be calculated in the classical limit in the bulk through
\begin{align}
    A = -T \log{Z} \overset{\text{classical}}{=} -iT S_c,
\end{align}
where $S_c$ is the on-shell periodic Euclidean time action. The on-shell field solutions only depend on the $z$ coordinate and are thus proportional to $V = i \beta V_2$, where $V_2$ is the area considered in coordinates $x_1$, $x_2$. This gives the free energy per surface area
\begin{align}
    \frac{A}{V_2} = \int_0^{z_h}{\text{d} z \sqrt{-g} \mathcal{L} + V^{-1} S_{\text{bdy}}}.
\end{align}

Only the free energy difference of the different solutions is needed. We therefore calculate the free energy contribution $A_{\text{fields}}$ from the scalar and electromagnetic fields and neglect the contribution from the Einstein-Hilbert term of the action. We do not need to account for the contribution from any backreaction on the metric following the previous argument. We first consider the trivial solution of
\begin{align}
   \frac{A_{\text{fields}}}{V_2} &= \int_0^{z_h}{\text{d} z \sqrt{-g} \mathcal{L} + V_2^{-1} S_{\text{bdy}}} \nonumber \\ &= -\int_0^{z_h}{\text{d} z (\frac{z}{L})^{-4} \frac{1}{4} F_{ab} F^{ab}} \nonumber \\ &= -\int_0^{z_h}{\text{d} z (\frac{z}{L})^{-4} \frac{1}{2} F^2_{zt} g^{zz} g^{tt}} \nonumber \\ &= \int_0^{z_h} \text{d} z \frac{\mu^2}{2z_h^2} \nonumber \\ &= z_h^{-1} \frac{\mu^2}{2} = \frac{4 \pi T}{3} \frac{\mu^2}{2}.
\end{align}
This agrees with the result from thermodynamics, with $\mu$ is the change in free energy for increasing the expectation value of the number of particles by one while keeping the temperature constant, as in
\begin{align}
    u = (\frac{\partial A}{\partial \braket{N}})_T.
\end{align}
This is easily shown by using $\rho = \braket{N}/V_2$ and we take Equation $~\ref{eqn:chem_trivial}$ to get the $N$ dependence of $\mu$ for constant $T$. The gravitational part of the free energy can be neglected since the derivative is set at constant temperature.

The free energy for the numerical solutions has been calculated and the result together with the analytical result is shown in Figure $~\ref{f:A}$.

\fig{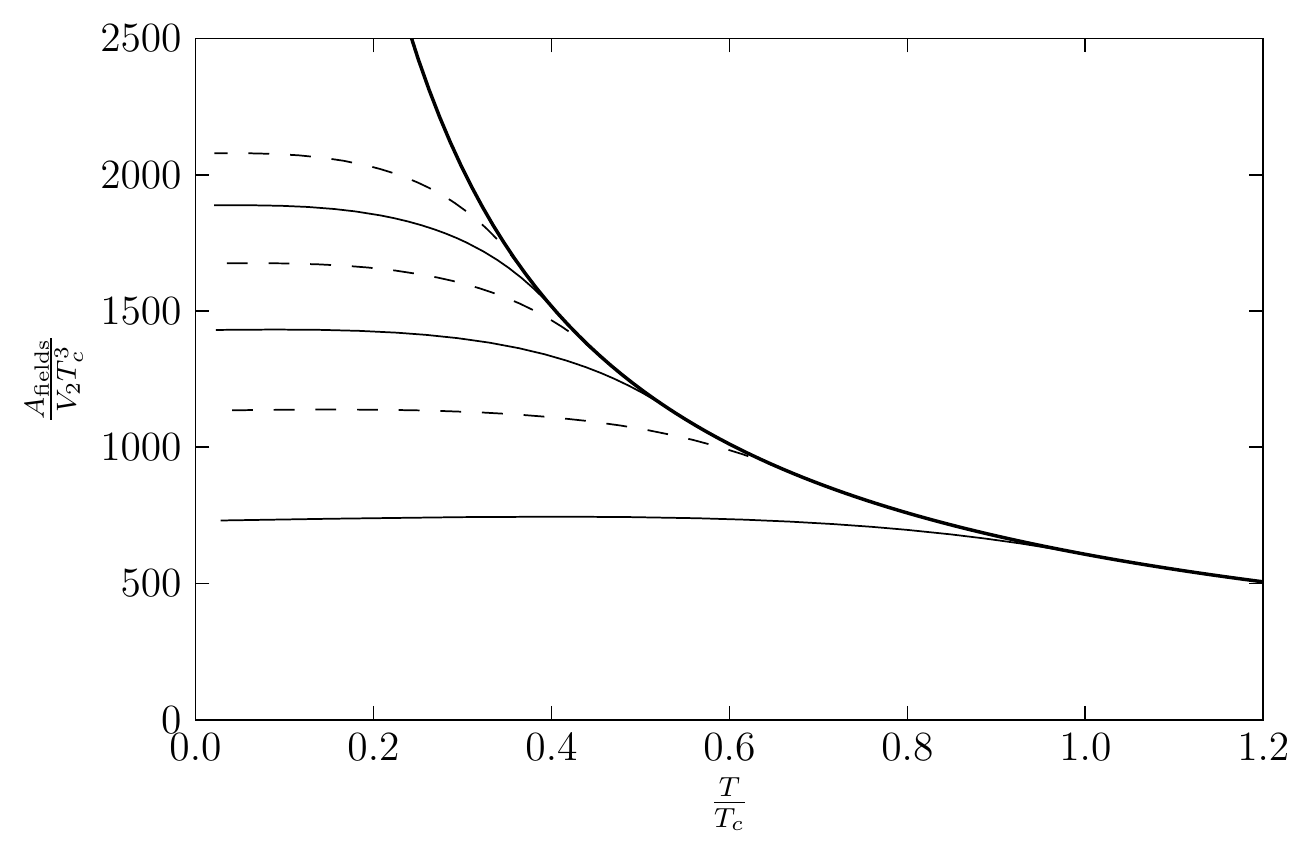}{The free energies of the different solutions are shown here for constant $\rho$ and varying $T$. The trivial solution is shown as the thick solid line. The other curves correspond to numerical solutions. The lowest one corresponds to the solution appearing at $T=T_c$ and the other roots follow in order.\label{f:A}}

It can there be seen that the trivial solution is the physical solution for temperatures above $T_c$ and that the solution appearing at the temperature $T_c$ is the physical solution for all lower temperatures. We will hereafter only work with these two solutions. The graph is in line with the phase transition being a second-order phase transition, though this has not been investigated mathematically.

\subsubsection{\label{sec:level1}Preliminary discussion of electrical conductivity}

The conductivity of a superconductor can easily be measured experimentally for a wide range of frequencies and it is therefore an interesting property to calculate from our model of a superconductor. The agreement in different frequency ranges tells us about similarities and differences between our model and the experimental superconductors.  

Let the conductivity $\sigma$ be the linear response function for the current density $J_x$ with applied electrical field $E_x$ as source
\begin{align}
    \sigma(\omega) = \frac{J_x(\omega)}{E_x(\omega)}.
\end{align}

From the Drude model, conductivity is also a complex quantity, with this current density-electric field relation suggesting that the imaginary part then represents the resistance to change of the electromagnetic field's charge carriers. The direction $x$ has been chosen for concreteness, but since we consider two-dimensional systems with rotational symmetry, we need only consider one direction. These functions of $\omega$ are the Fourier transforms of the time-dependent quantities. The current in the time domain can be obtained from the conductivity and the applied field through an inverse Fourier transform
\begin{align}
    J_x(t) = \int_{-\infty}^\infty{E_x(t - \tau)\sigma(\tau) \text{d} \tau}.
\end{align}
Causality implies that $\sigma(t) = 0$ for $\tau < 0$, since the current would otherwise depend on future values of the electrical field. Using this, the conductivity can be written 
\begin{align}
    \sigma(\omega) = \int_0^\infty{\sigma(\tau) \exp(i \tau \omega) \text{d} \tau}
\end{align}
and thus has an analytic extension to the upper half of the complex plane. Both the current, $J_x(t)$, and applied field, $E_x(t)$, are real quantities which make $\sigma(\tau)$ also real and thus $\text{Re}(\sigma(\omega))$ an even function and $\text{Im}(\sigma(\omega))$ odd. These properties of $\sigma(\omega)$ give the Kramers-Kronig relations
\begin{align}
    \text{Re}(\sigma(\omega)) &= \frac{2}{\pi} \int_0^\infty{\frac{\omega' \text{Im}(\sigma(\omega'))}{\omega{'}{^2} - \omega^2} \ \text{d} \omega'}
\end{align}
and
\begin{align} 
    \text{Im}(\sigma(\omega)) &= -\frac{2}{\pi} \int_0^\infty{\frac{\omega \text{Re}(\sigma(\omega'))}{\omega{'}{^2} - \omega^2} \ \text{d} \omega'},
\end{align}
which state that the real part of the conductivity uniquely determine the imaginary part and vice versa.

\section{\label{sec:level1}Determination and Assessment of the Holographic Conductivity}

We have discussed how the bulk equations of motion can be solved via an infinitesimal applied electrical field in the $x$ direction. The field was oscillating with a frequency $\omega$. The conductivity can now be calculated from the obtained current in the $x$ direction.
\begin{align}
    \sigma = \frac{J_x}{-\partial_t A_x} = -\frac{A_{x(1)}}{i \omega A_{x(0)} L}.
\end{align}
The electrical conductivity for the normal phase can now be found using the trivial solutions of Equation $~\ref{trivial}$. The boundary behavior is
\begin{align}
    A_{x(0)} = \lim_{z \rightarrow 0} A_x(z) = \exp(-\frac{i \pi z_h \omega}{6 \sqrt{3}}).
\end{align}
The conductivity is therefore 1 for all $\omega$ above $T_c$.

\begin{figure}
    \centering
    \includegraphics[width=0.50\textwidth]{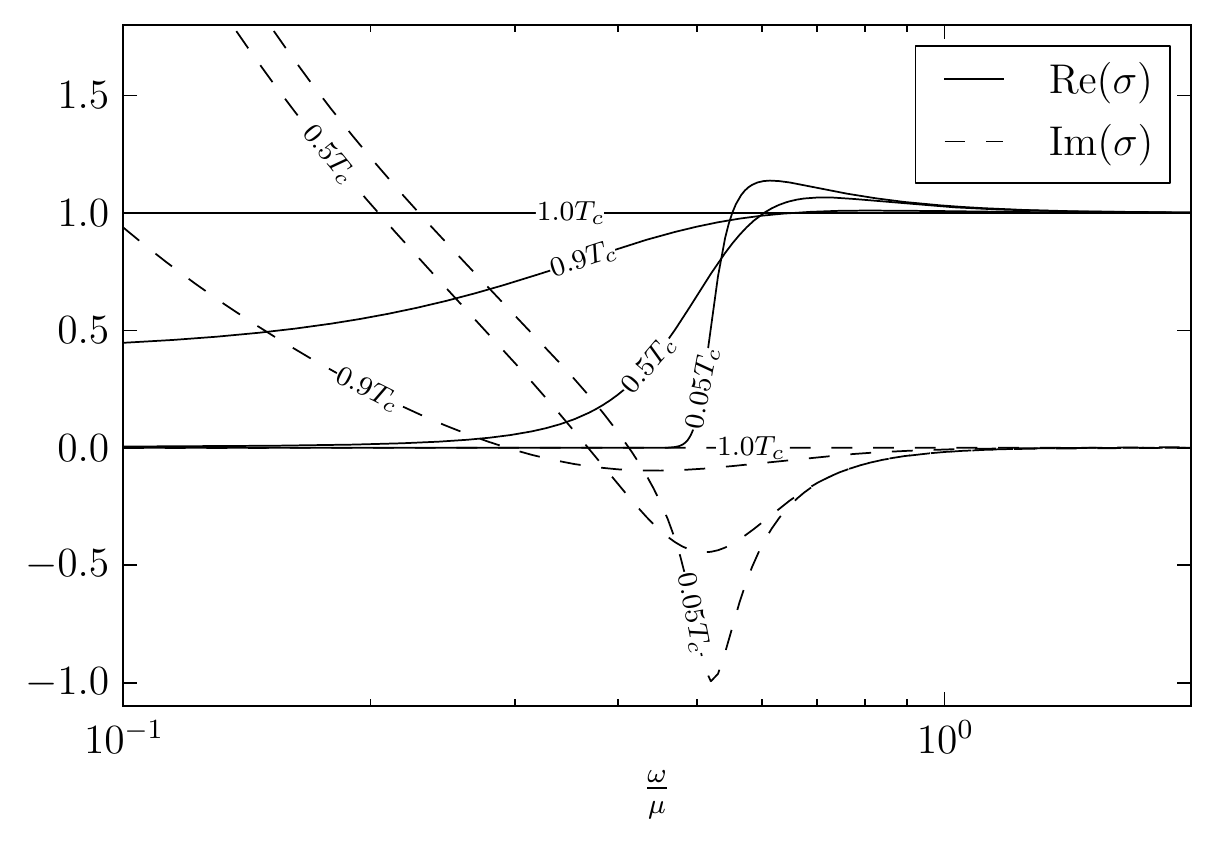}
    \caption{Real and imaginary part of the conductivity for different temperatures. $\rho$ is here constant.}
    \label{fig:cond}
\end{figure}

The conductivity below $T_c$ can be calculated the same way using the numerical solution. The result is seen in Figure $~\ref{fig:cond}$. The conductivity is lowered for low $\omega$ when the condensate forms. An energy gap $\Delta_{\text{gap}}$ forms and the conductivity for $\omega < \Delta_{\text{gap}}$ goes to 0 as the temperature is lowered. The superconductivity is not immediately evident from the obtained conductivity curves. There is, however, a delta function at $\omega = 0$ since translational invariance of the boundary theory has been assumed and the charged field $\psi$ has obtained a non-zero expectation value. The delta function can be seen through the Kramers-Kronig relations. A delta-function in the real part is equivalent to a pole in the imaginary part:
\begin{align}
    \text{Im}(\sigma(\omega)) &= -\frac{2}{\pi} \int_0^\infty{\frac{\omega \delta(\omega')}{\omega{'}{^2} - \omega^2} \ \text{d} \omega'} \nonumber \\ &= \frac{1}{\pi \omega}.
\end{align}
This pole is also visible in the figure.

The Kramers-Kronig relations and the independence of low energy properties for the high-frequency conductivity can be used to prove a sum rule for the conductivity\cite{PhysRev.109.1398}. The rule states that
\begin{align}
    \int_0^\infty{\text{Re}(\sigma(\omega)) \ \text{d} \omega} = C,
\end{align}
where $C$ is a constant depending on what system we are considering. This integral diverges in our case since $\sigma \rightarrow 1$ for high frequencies, so the rule must be modified. The proof uses that the imaginary part of $\sigma(\omega)$ becomes independent of low energy properties such as the temperature at high enough frequencies. That fact that the Kramers-Kronig relations apply to $\sigma(\omega)$ is also used. These two properties are also true for $\sigma(\omega) - 1$, so the rule can be modified into
\begin{align}
    \int_0^\infty{\text{Re}(\sigma(\omega) - 1) \ \text{d} \omega} = C.
\end{align}
The analytical solution above $T_c$ now gives $C=0$. This rule can now be used to verify that our numerics are accurate. The integral of $\text{Re}(\sigma(\omega) - 1)$ should vanish for all temperatures. It is then important to include the delta function at $\omega = 0$, which our numerics do not catch. We can, however, find the amplitude, $\Sigma_\delta$, of the delta function from the amplitude of the pole in the imaginary part of the conductivity. The sum of the integral of the continuous part of the conductivity, $\int{(\sigma_n - 1) \ \text{d} \omega}$, and $\Sigma_\delta$ should then equal 0 for all temperatures, as indicated by the red line in Figure $~\ref{f:sum2}$, with the results of each expression as a function of $T/T_c$ shown as $[T_c]$ on the vertical axis. 

A suitable cut-off frequency has been used, but since $\sigma(\omega)$ converges fast to 1, this is not a concern. No large discrepancies are observed. This is a strong check of the numerics, since the real part of the conductivity at all frequencies and temperatures and the amplitude of the pole all have to match up. The largest discrepancy is observed for low temperatures, which we can expect since the numerical integrator there has been observed to make smaller steps, indicating a numerically more difficult problem.

\begin{figure}
    \centering
    \includegraphics[width=0.50\textwidth]{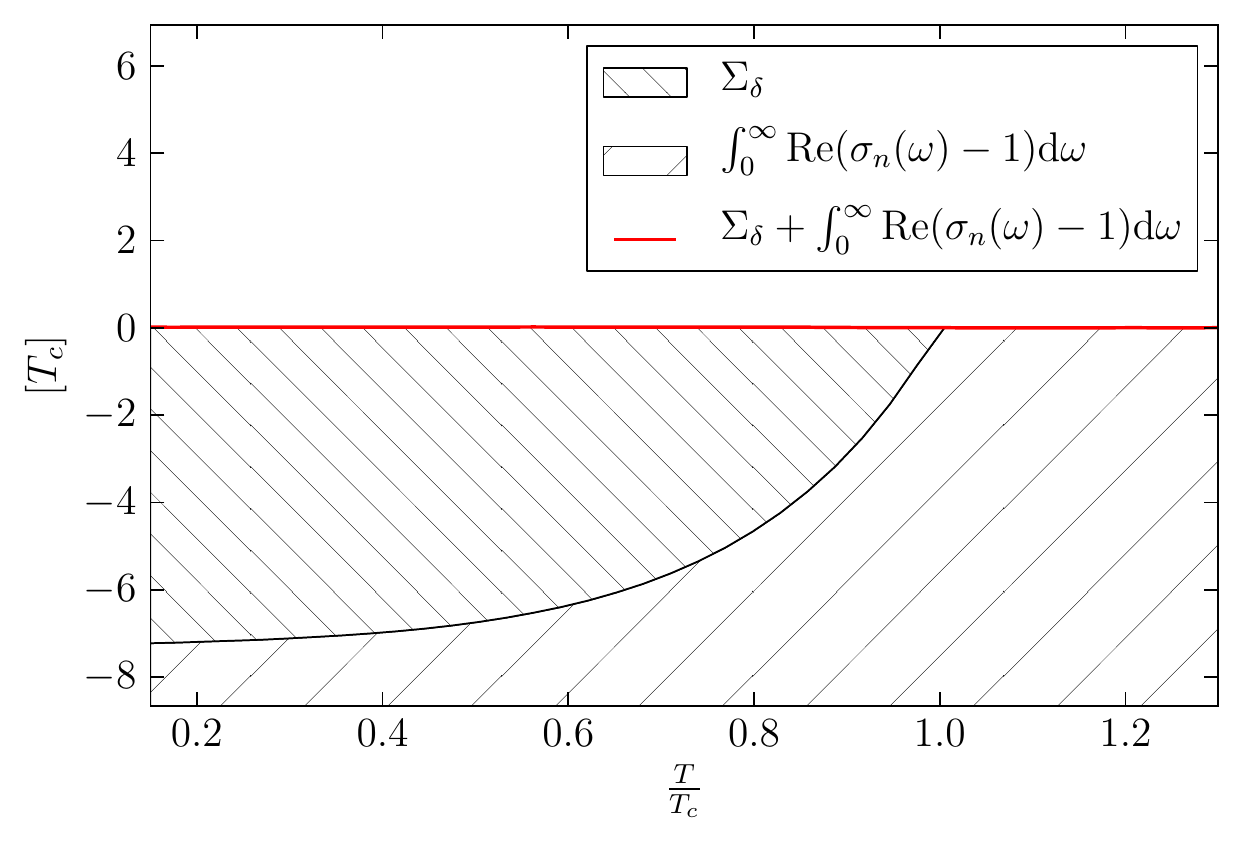}
    \caption{The two contributions to the integral in the modified Ferrell-Glover-Tinkham sum rule for different temperatures. The red line is expected to be precisely at 0 for perfect numerics.}
    \label{f:sum}
\end{figure}

\section{\label{sec:level1}Extending the Lagrangian}

This analysis makes several simplifications for the system at hand, the most prominent being the simple Lagrangian of Equation $~\ref{eqn:first_lagrangian}$ and that translational symmetry was assumed. This gives a boundary theory with a scalar field that condenses below a critical temperature, as expected for a superconductor. The conductivity shows both similarities and differences with that of high-$T_c$ superconductors. A delta function develops at $\omega = 0$ for $T < T_c$, giving infinite DC conductivity. An evident difference is that lack of a $\textit{Drude peak}$ at low frequencies, a Drude peak being an increase in conductivity for low frequencies in metals due to impurities that can be well modeled by the Drude model of conductivity\cite{drude}, hence the name. This model has been shown to agree with experiments on cuprates above $T_c$\cite{drudeFit}. The linear dependence found by H. L. Liu et al. is attributed to electron-phonon interactions. Electrons might not be a suitable type of excitation for these possibly strongly coupled systems and our model might capture interactions between some other type of excitations. The assumptions of temperature-independent Drude model parameters, $\rho$, $q^2$ and $m$, might then not be valid. The true temperature dependence of the scattering rate can easily be obtained directly from the Drude fit but this has not been done. This can then be compared with the linear experimental observations. The temperature dependence of the Drude model parameters can then be found. This might give important insights as to what types of excitations the Drude model in the cases we have studied describes with such high accuracy.

The Drude model is obtained by treating the charge carriers classically. They are expected to obey the differential equation
\begin{align}
    \frac{\text{d} v}{\text{d} t}
 = \frac{q}{m} E - \frac{1}{\tau} v,
 \end{align}
where $qE$ is the electric force, $m$ is the charge carrier mass, $q$ is the charge, and $\tau$ is the average time between collisions. The last term is a drag force supposed to model the collisions slowing the charge carriers down. Solving for this harmonic, $E = E_0 \exp(-i \omega t)$, gives
\begin{align}
     v = \frac{\tau qE_0}{m(1 - i \omega t)} \exp(-i \omega t).
\end{align}
This gives the conductivity
\begin{align}
    \sigma(\omega) = \frac{J(\omega)}{E(\omega)}= \frac{v(\omega)q \rho}{E(\omega)} = \frac{\tau \rho q^2}{m(1 - i \omega \tau)} = \frac{\sigma_0}{1 - i \tau \omega},
\end{align}
where $\rho$ is the density of charge carriers of some charge $q$. From this definition, we have
\begin{align}
    \frac{\sigma_0}{\tau} = \frac{\rho q^2}{m}.
\end{align}
Different generalizations of the standard Lagrangian of Equation $~\ref{eqn:first_lagrangian}$ have been studied. Higher order corrections using $\psi$ will not make a difference above $T_c$, so a Drude peak cannot be obtained using them. T. Wenger studied these extensions\cite{Wenger2012}, and this reassessment will use his added higher order Maxwell term, resulting in
\begin{align}
    \mathcal{L} &= \frac{1}{2 \kappa} (R - 2 \Lambda) - \frac{1}{4} f_{ab} F^{ab} \nonumber \\ &- m^2 \psi \overline{\psi} - D_a \psi \overline{D^a \psi} \nonumber \\ &+ \alpha_2 F^a_b F^b_c F^c_d F^d_a,
\end{align}
which gives equations of motion different from our original Equation $~\ref{eqm1}$, $~\ref{eqm2}$, and $~\ref{eqm3}$. Using SymPy, $z_h = 1$ and $q = 1$ instead yield
\begin{align}
    (-4z^3 + 2z^2 \phi^2 + 4) &\psi  \nonumber \\ + (2z^7 + 2z^4 - 4z) &\psi' \nonumber \\ + (2z^8 - 4z^5 + 2z^2) &\psi'' = 0
\end{align}
for $\psi$,
\begin{align}
    (z^5 - z^2) &\phi''  \nonumber \\ + (24\alpha_2 z^9 \phi'' - 24\alpha_2 z^6 \phi'')&\phi{'}{^2} \nonumber \\ + (32\alpha_2z^8 - 32 \alpha_2 z^6) &\phi{'}{^3}  \nonumber \\+ 2 \psi^2 &\phi = 0
\end{align}
for $\phi$, and
\begin{align}
    (8\alpha_2 z^{12} \phi{'}{^2} - 16 \alpha_2 z^9 \phi{'}{^2} + 8\alpha_2z^6 \phi{'}{^2} + z^8 - 2z^5 + z^2) &A_x''  \nonumber \\+  (8\alpha_2 \omega^2 z^6 \phi{'}{^2} + \omega^2 z^2 + 2z^3 \psi^2 - 2 \psi^2) &A_x  \nonumber \\ + (16\alpha_2 z^{12} \phi' \phi'' + 56 \alpha_2 z^{11} \phi{'}{^2} - 32 \alpha_2 z^9 \phi' \phi'' - 88\alpha_2z^8 &\phi{'}{^2} + \nonumber \\ 16\alpha_2 z^6 \phi' \phi'' + 32 \alpha_2 z^6 \phi{'}{^2} + 3z^7 - 3z^4)&A_x' = 0
\end{align}
for $A_x$. 

\begin{figure}
    \centering
    \includegraphics[width=0.5\textwidth]{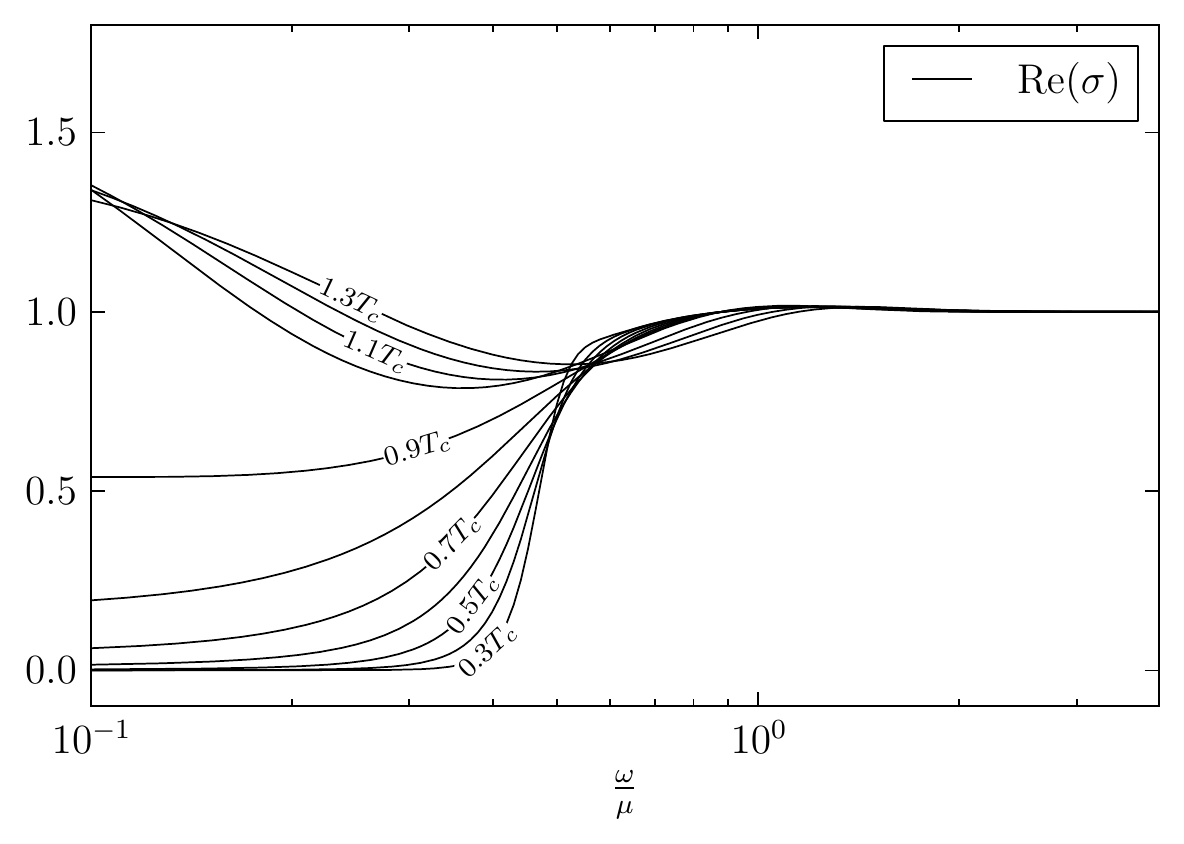}
    \caption{Real part of the conductivity for different temperatures using the extended Lagrangian with $\alpha_2=0.01L^4$. $\rho$ is here constant.}
    \label{f:cond_a2_a1}
\end{figure}

The exponents of the horizon and boundary are identical because the higher order term vanishes faster both at the horizon and the boundary. The analytical solution of the equations of motion found above $T_c$ for the original Lagrangian is no longer valid and a numerical solution must now be used above $T_c$ as well. The result of the conductivity calculated with this extended model can be seen in Figure $~\ref{f:cond_a2_a1}$, where small $\alpha_2$ has been used.

\begin{figure}
    \centering
    \includegraphics[width=0.50\textwidth]{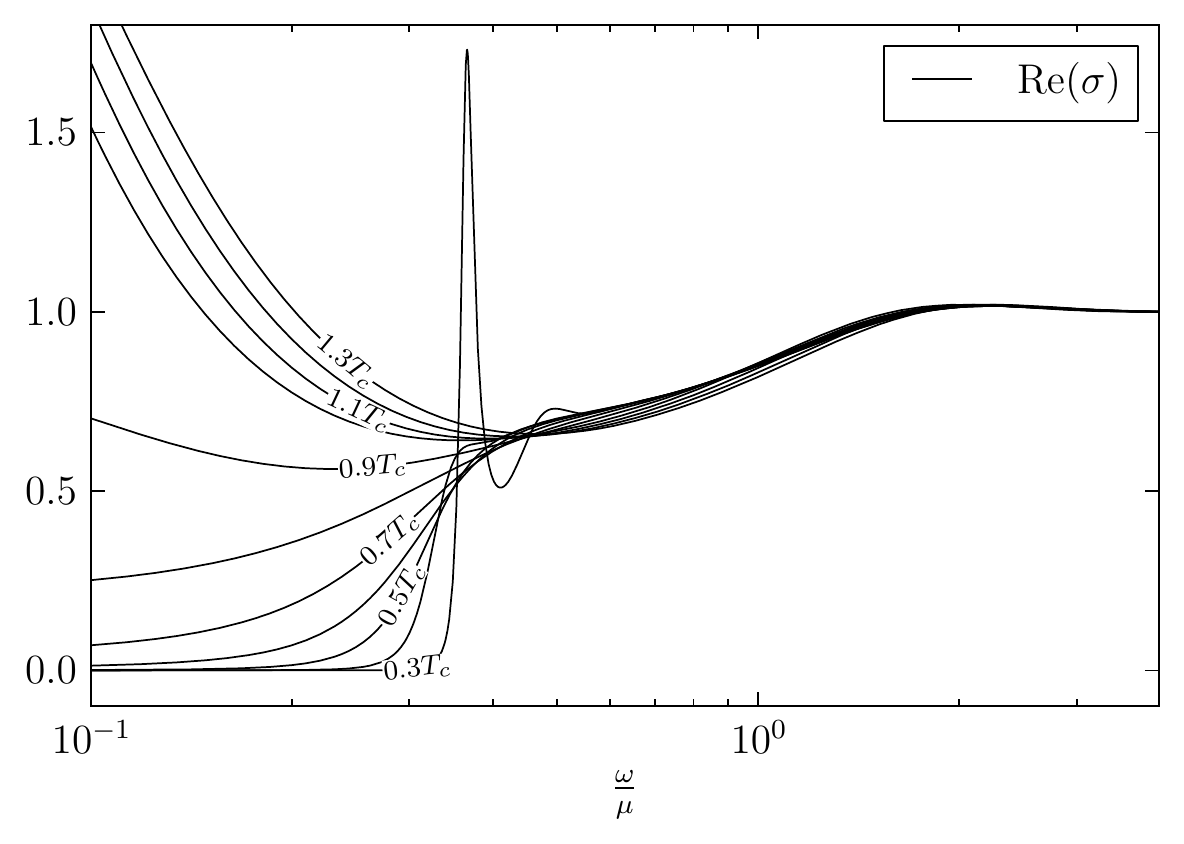}
    \caption{Real part of the conductivity for different temperatures using the extended Lagrangian with $\alpha_2=0.1L^4$. $\rho$ is here constant.}
    \label{f:cond_a2_2}
\end{figure}

The higher order term introduces a perturbation in the low frequency conductivity. The gap still appears at the same position, $\omega_{\text{gap}} \approx \mu/2$. The behavior around the transition from $\sigma = 0$ to $\sigma = 1$ is changed. The earlier increase in conductivity for $\omega$ slightly larger than $\omega_{\text{gap}}$ has been changed. The increase is lower and at higher $\omega$. The conductivity for a slightly larger $\alpha_2$ is seen in Figure $~\ref{f:cond_a2_2}$. Here, the change around the transition is more pronounced. A peak develops for low temperatures and a second peak is also seen to develop. More peaks develop for higher $\alpha+2$ and lower $T$. These peaks seem to approach delta functions when the temperature is lowered further. The peaks are not believed to be a numerical error, since their contribution is needed for the sum rule to be satisfied.

The transition to higher conductivity is now seen to happen in two steps. First, the conductivity increases to $\sigma \approx 0.7$ at $\omega$ slightly lower than $\omega_{\text{gap}}$. Now this transition also happens for higher temperatures, the conductivity approaches 0.7 from above. Secondly, the conductivity increases to reach 1 at $\omega \approx 2$; this second part of the transition seems to be mostly independent of temperature.

The conductivity behavior above $T_c$ resembles that of the Drude model, however. A fit of the Drude parameters $\sigma_0$ and $\tau$ can be made to see how well this model agrees with our conductivity. One point of the complex conductivity is enough to obtain both of these real parameters. The drude model conductivity approaches 0 at high frequencies whereas the conductivity of our holographic model approaches 1. We can thus not expect the model to work well for high frequencies. We make the Drude fit by taking the value of the conductivity in the limit $\omega \rightarrow 0$. First, obtain $\sigma_0$ through
\begin{align}
\sigma_0 = \lim_{\omega \rightarrow 0} \sigma(\omega).
\end{align}
Use this to obtain $\tau$:
\begin{align}
    \lim_{\omega \rightarrow 0} \frac{\text{Im}(\sigma)}{\omega \sigma_0} = \lim_{\omega \rightarrow 0} \frac{\tau}{1 + \tau^2 \omega^2} = \tau.
\end{align}
A fit using these obtained parameters is shown in Figure \ref{f:drude}. The fit agrees well for low frequencies but a difference of 1 in the real part appears as the Drude model conductivity approaches 0 and the holographic conductivity approaches 1. This agreement might not be very impressive, but the Drude peak grows when $\alpha_2$ is increased and relative error vanishes in the limit of large $\alpha_2$. See Figure \ref{f:drude2} for a Drude fit with a higher $\alpha_2$. The low frequency conductivity is here much larger but the error is still of order 1 and only appears at higher frequencies where the holographic conductivity approaches 1.

\fig{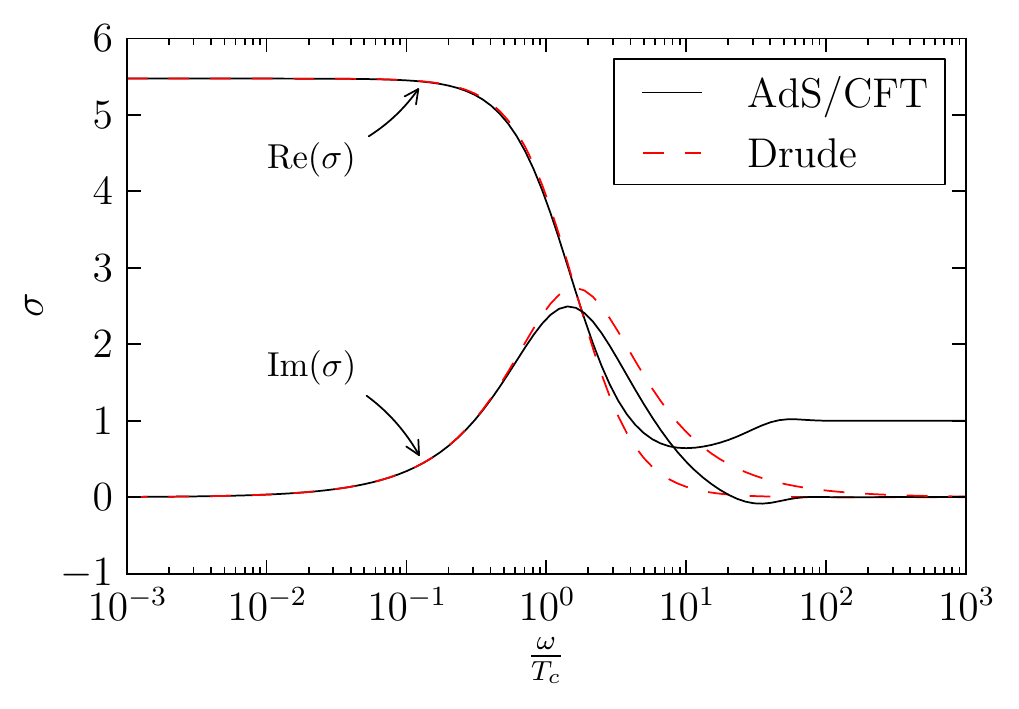}{Conductivity for $\alpha_2=0.1L^4$ and $T=T_c$ together with Drude model fit.\label{f:drude}}

\fig{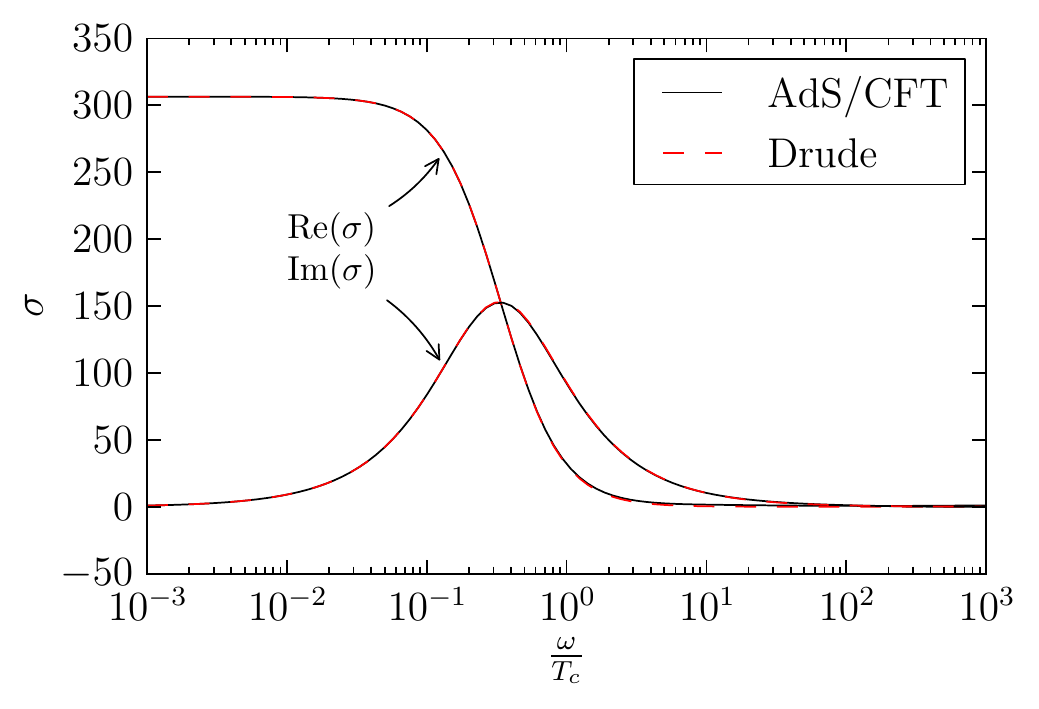}{Conductivity for $\alpha_2=10L^4$ and $T=T_c$ together with Drude model fit.\label{f:drude2}}

This agreement then motivates an investigation of the Drude parameters dependence on system parameters. Figure \ref{f:drudeVar} shows the dependence of the parameters on the strength of the higher order term, $\alpha_2$. A power-law was fit to both $\sigma_0$ and $\tau$ at high $\alpha_2$. $\sigma_0$ clearly approaches a linear dependence on $\alpha_2$. $\tau$ did not permit a power-law description.\\

\fig{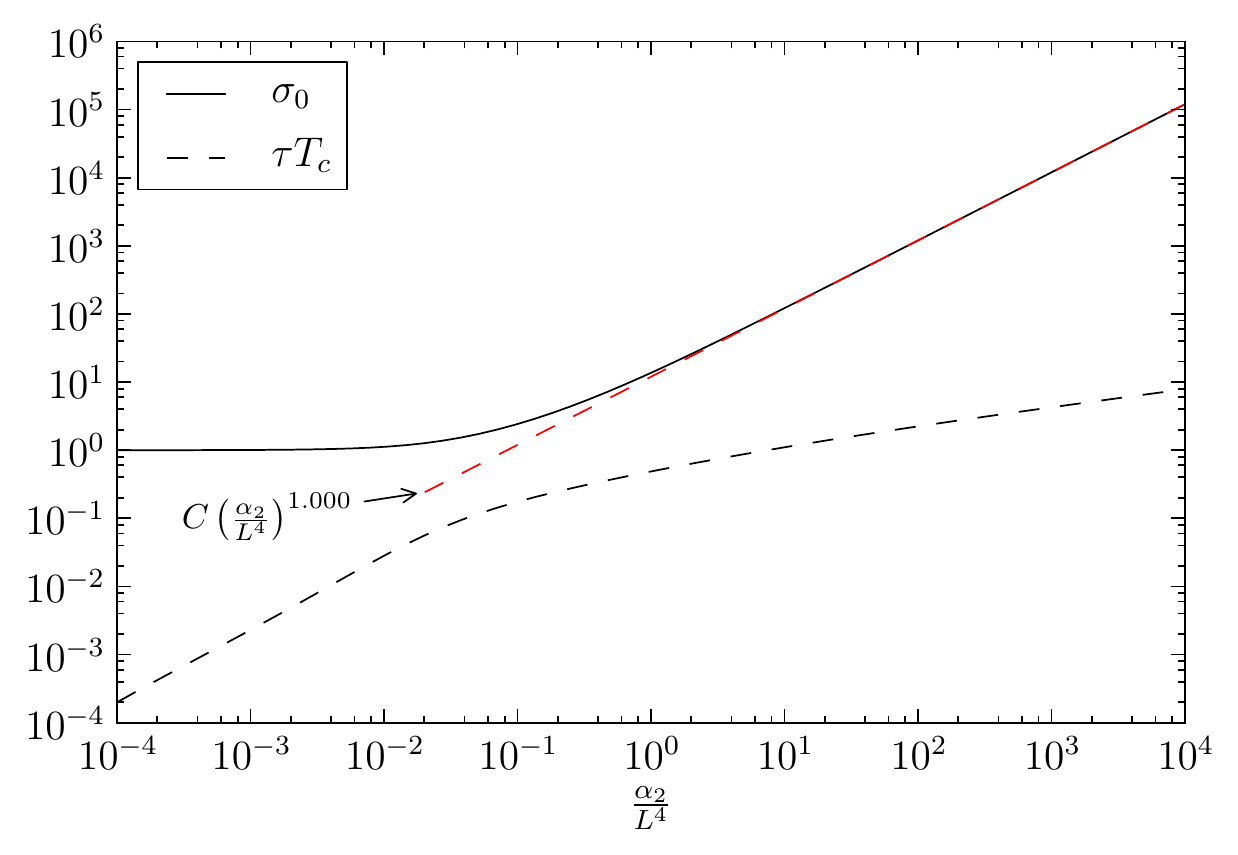}{Drude parameters as functions of $\alpha_2$ at $T=2T_c$.\label{f:drudeVar}}

\begin{figure}[H]
    \centering
    \includegraphics[width=0.50\textwidth]{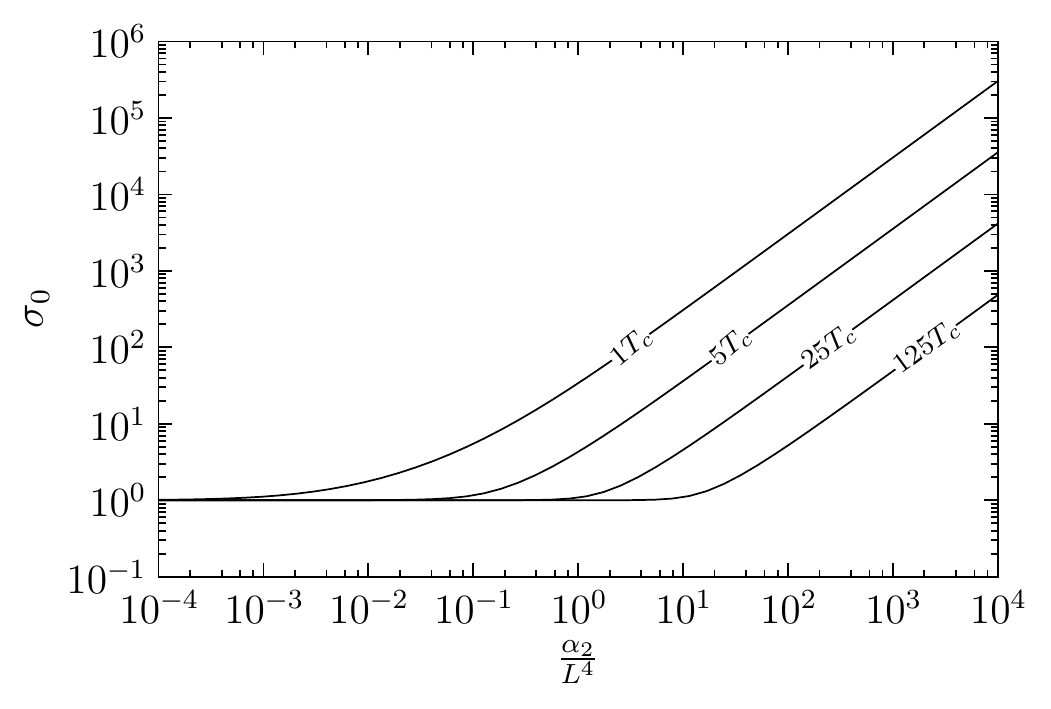}
    \caption{$\omega\rightarrow0$ limit of the conductivity as a function of $\alpha_2$ for different temperatures.}
    \label{f:drudeVarT}
\end{figure}

Now we investigate the temperature dependence of $\sigma_0$ to try to see how the coefficient $C$ in the linear dependence on $\alpha_2$ depends on temperature. Figure $\ref{f:drudeVarT}$ shows the dependence of $\sigma_0$ on both $\alpha_2$ and $T$. The coefficient $C$ also seems to follow a power-law since the separations between the curves corresponding to different temperatures are equal. A plot of the temperature dependence of $C$ is seen in Figure \ref{f:Cdep}. A formula for $\sigma_0$ valid for large $\alpha_2T_c^{-4/3}/(T^{-1/3}L)^4$ can according to these results be expressed as
\begin{equation}
 \sigma_0=C\frac{\alpha_2}{L^4}\left(\frac{T}{T_c}\right)^{-4/3}.
\end{equation}
This is valid for a large range of values of $\alpha_2$ and $T$.

\fig{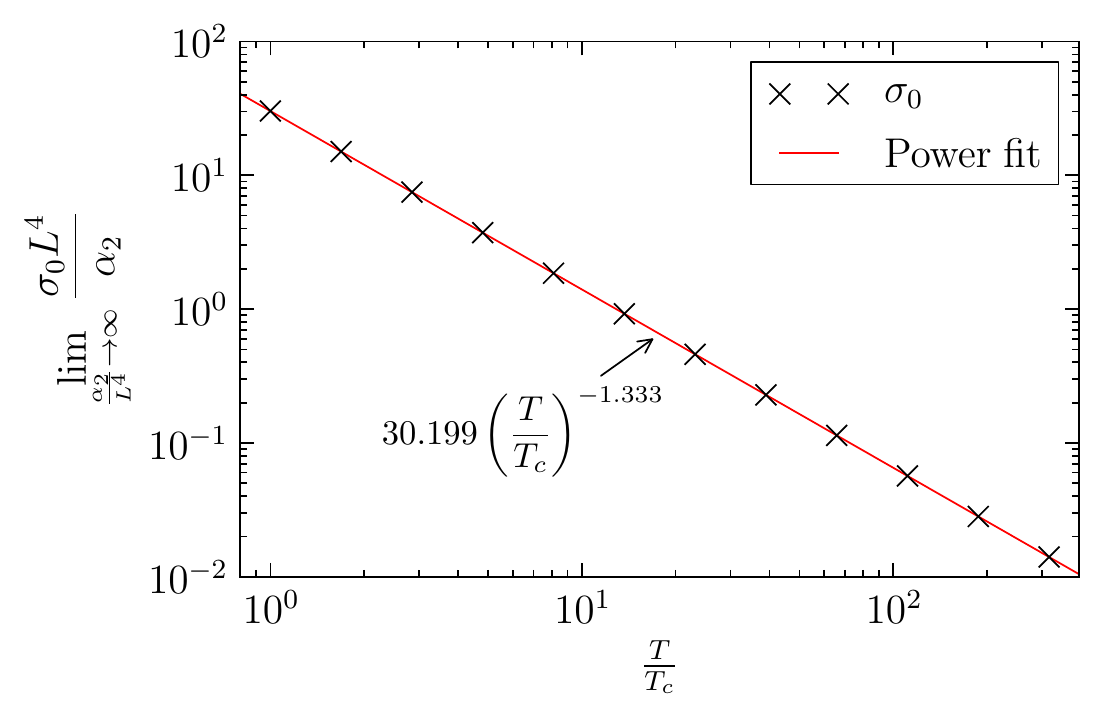}{$\sigma_0L^4/\alpha_2$ as a function of temperature for large $\alpha_2$.\label{f:Cdep}}

A test of the numerics through a consistency check has also been performed, with the results shown in Figure $~\ref{f:sum2}$. There is a substantial error at low temperatures, but it is decreased by increasing the number of samples for the numerical integral, so it is believed to originate from the very sharp peaks, approaching delta functions, appearing at low temperatures making numerical integration difficult.

\fig{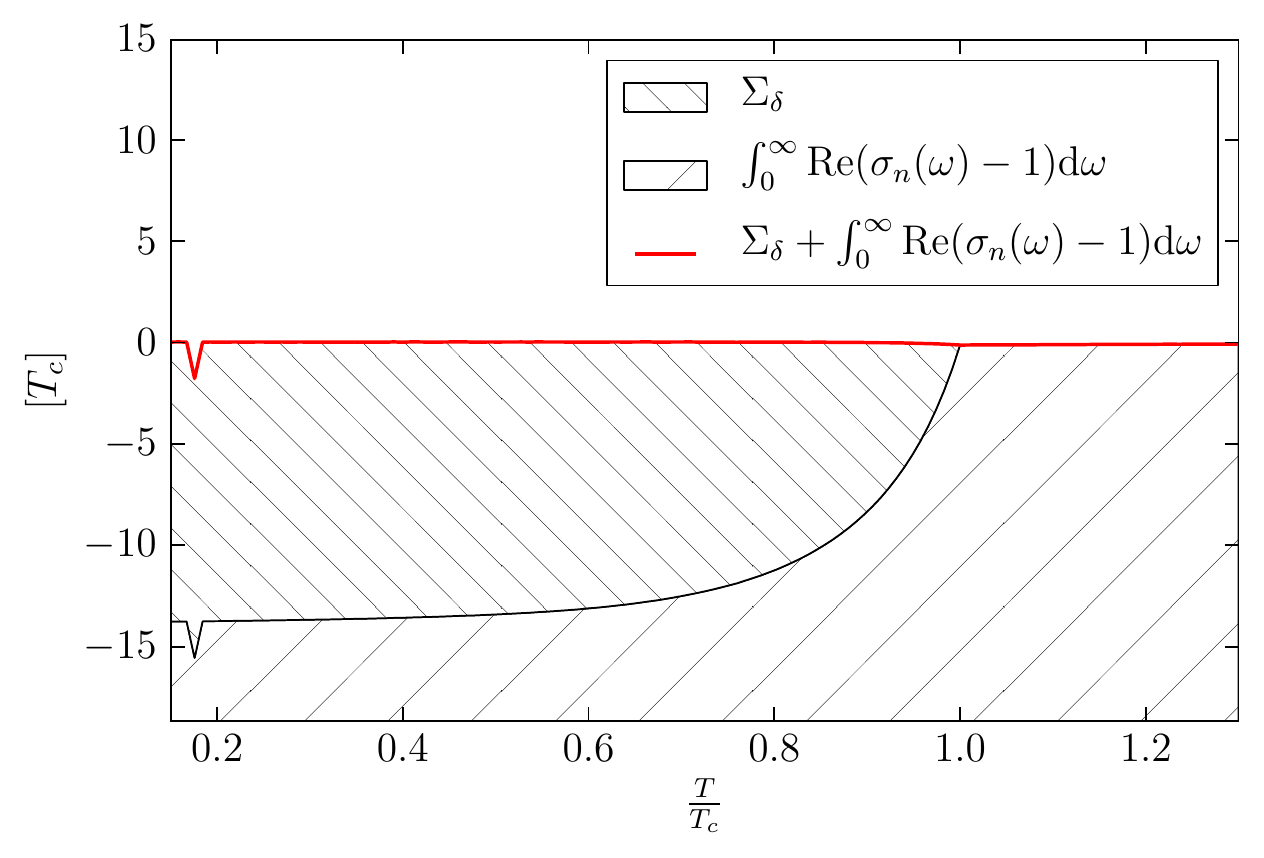}{The two contributions to the integral in the modified Ferrell-Glover-Tinkham sum rule for different temperatures. The red line is expected to be precisely at 0 for perfect numerics\label{f:sum2}. The small bump at about $T=0.2T_c$ is due to the numerical integration missing the very sharp conductivity peak developing at low temperature.}

\section{\label{sec:level1}Conclusions}
Through the modeling work of researchers like G. T. Horowitz and P. Säterskog, the AdS/CFT correspondence can be used to model a high-$T_c$ superconductor both below and above $T_c$. This was initially done using the simplest possible Lagrangian and then calculating the frequency-dependent conductivity. A higher order term was added to the Lagrangian and a conductivity peak was obtained at low frequencies. The Drude model describes this peak very well within certain limits. The behavior of the Drude model parameter $\sigma_0$ was investigated in certain limits. The extended Lagrangian seems to give an effective description of the lattice introduced by G. T. Horowitz and J. E. Santos\cite{horowitz}, but further investigations are of interest.

Overall, the topic is in need of a more phenomenological perspective regarding real-world experimental data. The ramifications of Equation $~\ref{eqn:chem_trivial}$, for example, which relates the chemical potential and charge density for the non-superconducting solution, could be further examined. This has been done by S. A. Hartnoll\cite{Hartnoll2009}, but this could be replicated. A thorough treatment of the differential equations would also be of interest. All curves in Figure $~\ref{f:O}$ approach the same value. The differential equation has been numerically investigated and it has the same boundary behavior, regardless of initial conditions, so it seems possible to find a mathematical explanation for the behavior by looking at the boundary behavior of the equations.

As P. Säterskog\cite{Saterskog13} suggests, the results from the extended Lagrangian could be investigated further. A physical interpretation of the conductivity peaks for low temperatures is required for an understanding of the extension to the Lagrangian. The temperature dependence of the scattering rate $1/\tau$ can be found and compared to experiment. A linear temperature dependence would agree with experiments and thus indicate that the extended Lagrangian truly captures the physics giving rise to the conductivity peak. A more thorough comparison with the results of G. T. Horowitz$\cite{horowitz}$ is of interest, since we could hope to have found an effective description of their system. The power law behavior of the conductivity at intermediate frequencies could also further be investigated.
\section{\label{sec:level1}Acknowledgements}
The author would like to gratefully acknowledge Professor David Kaplan of Johns Hopkins University for his recommendations and research overview.
\newline
\bibliography{main}
\end{document}